\documentclass[]{aa}  

\usepackage{graphicx}
\usepackage{txfonts}
\begin{document} 

\title{Close-by planets and flares in their host stars}
   \authorrunning{Lanza et al.}
   \subtitle{}
\author{A.~F.~Lanza}
\institute{INAF-Osservatorio Astrofisico di Catania, Via S.~Sofia, 78 - I-95123 Catania, Italy\\
              \email{nuccio.lanza@oact.inaf.it} }


   \date{Received 21 June 2017 ; accepted 23 october 2017 }

\abstract{The interaction between the magnetic fields of late-type stars and their close-by planets {{  may}} produce stellar flares as observed in active binary systems. {{  However, in spite of several claims,  conclusive evidence is still lacking. }}}{We estimate the magnetic energy available in the interaction  using analytical models {{  to provide an upper bound to the expected flare energy}}.}{We investigate three different mechanisms leading to magnetic energy release. The first two can release an  energy up to $(0.2-1.2)\, B^{2}_{0}R^{3}/\mu$, where $B_{0}$ is the surface field of the star, $R$ its radius, and $\mu$ the magnetic permeability of the plasma. They operate in young active stars whose coronae have closed magnetic field lines up to the distance of their close-by planets that can trigger the  energy release. The third mechanism operates in weakly or moderately active stars having a coronal field with predominantly open field lines at the distance of their planets. The released energy  is of the order of $(0.002-0.1)\, B^{2}_{0}R^{3}/\mu$ and depends on the ratio of the planetary to the stellar fields thus allowing an indirect measurement of the former when the latter is known.}{We compute the released energy for different separations of the planet and different stellar parameters finding the conditions for the  operation of the proposed mechanisms. An application to eight selected systems is presented.}{The computed  energies and dissipation timescales are in agreement with flare observations in the eccentric system HD~17156 and in the circular systems HD~189733 and HD~179949. This kind of star-planet interaction can be  unambiguously identified by the higher flaring frequency expected close to periastron in eccentric systems.} 

  \maketitle
%

\section{Introduction}
\label{intro}
Late-type main-sequence stars have surface magnetic fields that extend into their outer atmospheres and stellar winds. They can be studied in detail in our Sun \citep[e.g.][]{Priest84} and are detected in distant stars by means of spectropolarimetric techniques \citep[][]{DonatiLandstreet09}. In that case, they are generally studied by detecting their effects, such as cool spots in their photospheres, non-radiative heating of chromospheres and coronae, and transient energy release events such as flares. 

Many planets orbiting at distances between 0.02 and 0.15~AU around late-type stars have been detected by means of  stellar radial velocity measurements or through their transits across the discs of their stars. Giant planets with orbital semimajor axis  $\la 0.1$~AU and masses comparable with that of Jupiter are called Hot Jupiters and are efficiently detected by those methods. Those with an orbital period longer than $7-10$ days generally have eccentric orbits, some of which with eccentricities larger than 0.6-0.7 \citep{UdrySantos07}. At those close separations, these planets orbit inside the magnetic fields of the coronae of their host stars or in the accelerating regions of their stellar winds.  Therefore, a fundamental question is how they interact with a stellar corona or a wind and how the energy of the stellar magnetic field is perturbed by their presence. 

\citet{RubensteinSchaefer00} conjectured that close-by planets may induce large flares in late-type stars by a mechanism similar to that observed in close binary systems with late-type components, notably RS CVn and Algols. Their proposal was motivated by the observations of super-flares with energies up to $10^{26}-10^{31}$~J in some otherwise normal solar-type stars \citep{Schaeferetal00}. Indeed, close binary systems with eccentric orbits provided evidence for a higher frequency of flares at periastron. One of the best examples is the T~Tauri binary system V773~Tauri \citep{Massietal02}. Its very young and highly magnetically active stars show giant magnetic loops or coronal helmet streamers that extend up to $\sim 20-30$ stellar radii in radio VLBI maps \citep{Massietal08} and interact strongly  producing intense flares preferentially at periastron. 

In the case of hot Jupiters on eccentric orbits, one would expect  a similar phenomenon. {  Nevertheless, the present evidence is not conclusive and still strongly debated owing to the limited number of observations and the lower energy and frequency of the flares. \citet{Maggioetal15} observed a flare in HD~17156 at the periastron during a simultaneous X-ray and Ca II H\&K monitoring campaign, while no similar activity was observed at other orbital phases. However, the limited number of observations did not allow to demonstrate the repeatability of the flaring at that orbital phase.} A  campaign conducted on HD 80606, which in principle should display a stronger effect, failed to provide evidence of flaring activity  close to periastron  \citep{Figueiraetal16}. In HD~189733, recurrent flares following the egress of the hot Jupiter from the occultation have been observed. In this system the orbit is circular, but the preferential orbital phase of the flares stimulated conjectures on the star-planet interaction mechanism(s) that could produce them \citep{Pillitterietal11,Pillitterietal14,Pillitterietal15}. {  However, flares can occur also during planetary transits \citep[e.g.][]{Klocovaetal17,Cauleyetal17} and their higher probability of occurrence close to the occultation of the planet needs more observations to be confirmed. }

For a better understanding of these phenomena, an estimate of the energy made available to produce flares in the coronae of the host stars during star-planet interaction is required. It has been obtained by means of numerical magnetohydrodynamic (MHD) models tailored on the parameters of specific star-planet systems, thus requiring extrapolations to other cases. Moreover,  MHD models generally assume a circular orbit and a stationary regime to simplify the numerical set-up  and the computations \citep[e.g.,][]{Cohenetal09,Cohenetal11,Strugareketal15,Strugarek16}. Therefore, it is useful to introduce a general analytic formalism to estimate the impact of close-in planets on the energy of the coronal fields of their stars. For the sake of simplicity, two scenarios will be considered: a) a planet orbiting inside the closed corona of its star; this is the case of close-by planets orbiting young T~Tauri or zero-age main-sequence stars that are highly active and rapidly rotating with closed coronal loops extending up to tens of stellar radii; b) a planet orbiting in the accelerating region of the wind of its star, where the magnetic field lines are open and radially combed by the wind flow; this applies to older stars with a moderate or low level of activity because their magnetic fields are not strong enough to confine the hot coronal plasma up to the distance of their planets. Our analytical models will allow to treat the case of close-in planets with both circular and eccentric orbits for an application to the above considered cases. 

\section{Models}
\subsection{The magnetic field of the stellar coronae}
\label{star_field_intro}

Models of the magnetic fields of the outer atmosphere of the Sun, applicable also to other late-type stars, are reviewed by, e.g., \citet{Wiegelmannetal15}. The parameter $\beta = 2\mu p/B^{2}$, i.e., the ratio of the thermal pressure $p$ to the magnetic pressure $B^{2}/2{\mu}$, where $\vec B$ is the magnetic field and ${\mu}$ the magnetic permeability of the plasma, can be used to characterize the MHD regime of the coronal plasma. In the lower corona, the magnetic pressure dominates over the thermal pressure ($\beta < 1$) and the field lines are typically closed, while in the outer regions the pressure of the plasma prevails opening up the field lines and accelerating the plasma to form the stellar wind. 

The dependence of the parameter $\beta$ on the radial distance $r$ from the centre of the star was studied in Sect.~2.1 of \citet{Lanza12} by a simple magnetohydrostatic model and we refer the reader to that work for details. Assuming an isothermal corona with a temperature $T$, that model gives $\beta$  on the equatorial plane, where the effect of the centrifugal force is maximum, as:
\begin{equation}
\beta(r) = 2 \mu p(R) B^{-2}(r) \exp \left\{ -\frac{R}{H_{0}} \left[ \left( 1- \frac{R}{r} \right) - \epsilon_{\rm rot} \left(1 - \frac{r^{2}}{R^{2}} \right) \right] \right\},
\label{beta_vs_radius}
\end{equation}
where $R$ is the radius of the star, $p(R)$ the plasma pressure at the surface of the star, $H_{0} = 5.1 \times 10^{7} (T/10^{6}) (R/R_{\odot})^{2} (M/M_{\odot})^{-1}$~m the pressure scale height with $T$ in K, and $\epsilon_{\rm rot} = \Omega^{2} R^{3}/(2GM)$ the ratio of the centrifugal to the gravitation potential on the equator of the star; {  here  $\Omega$ is the angular velocity of the stellar rotation, $M$ the mass of the star, and $G$ the gravitation constant.} The surface pressure $p(R) = 2 k_{\rm B} n_{\rm e} T$, where $k_{\rm B}$ is the Boltzmann constant and $n_{\rm e}$ the electron density at the base of the corona, and we assume a completely ionized hydrogen plasma. The dependence of the magnetic field strength $B(r)$ on the radial distance $r$ is a function of the adopted magnetic field model and will be specified below. For example, in  the case of a potential dipole field: $B(r) = B(R) (r/R)^{-3}$. 

In the closed corona, we assume that the magnetic pressure dominates over the gravity, the  plasma pressure, and the kinetic energy density $\frac{1}{2}\rho v^{2}$, where $\rho$ is the density of the plasma and $v$ its velocity. This is equivalent to $\beta \ll 1$ and $v \ll v_{\rm A} \equiv B/\sqrt{{\mu} \rho}$, where $v_{\rm A}$ is the Alfven velocity. Under these hypotheses, we can assume a magnetohydrostatic force-free model for the field \citep{Priest84,Wiegelmannetal15}, i.e. 
\begin{equation}
\nabla \times {\vec B} = \alpha {\vec B}, 
\label{fff_def}
\end{equation}
where the force-free parameter $\alpha$ is constant along each field line as follows from $\nabla \cdot {\vec B} = 0$. In general, $\alpha$ changes from one field line to the next  in non-linear force-free field models, while the case of constant $\alpha$ is referred to as that of linear force-free models. 

{  A basic constraint on the evolution of the magnetic field is imposed by the conservation of the magnetic helicity that is the volume integral of ${\vec A} \cdot {\vec B}$, where $\vec A$ is the vector potential of the magnetic field, i.e., ${\vec B} = \nabla \times {\vec A}$ \citep{Woltjer58,Priest84}. For  a  field confined within a closed volume, this definition is gauge-invariant, but this is not the case for stellar magnetic fields that cross the photosphere. The gauge invariance is restored by introducing the relative magnetic helicity defined as the difference between the magnetic helicity of the given field and that of the potential field with the same boundary conditions \citep{BergerField84,Berger85}. The dissipation of the relative magnetic helicity in a stellar corona is extremely slow, so it can be considered constant during the field evolution, even in the presence of magnetic reconnection \citep{Berger84,Berger85,HeyvaertsPriest84}. }

In a previous work, we discussed the  linear force-free field configurations suitable to describe the coronae of stars with close-by planets \citep{Lanza09}. In the present work, we want to consider the case of non-linear force-free fields; moreover, we assume that the planet has its own magnetic field. 

\subsection{The case of young and active stars}

In young, rapidly rotating stars, the closed corona extends up to several stellar radii. The observations of prominence-like structures, that is plasma condensations with temperature  $\approx 10^{4}$~K producing absorption features moving across the H$\alpha$ line profile in AB~Doradus, show that closed loops capable of confining these relatively cool and dense structures up to $\approx 10$ stellar radii exist in that young star \citep{CameronRobinson89a,CameronRobinson89b,Cameronetal90}. In addition to the case of V773~Tau mentioned in Sect.~\ref{intro}, this result supports the assumption of a corona with closed magnetic field lines up to the  distance of close-by planets in the range between 3 and 15 stellar radii. In Sects.~\ref{star_field} and~\ref{energy_star_field}, we focus on such young and active stars assuming that $\beta < 1$ up to the distance of the planets, while we shall consider the case of stars with a weaker field in Sect.~\ref{potential_fields}. 

\subsubsection{Magnetic energy available when the field can reach an open configuration}
\label{star_field}

{  A first mechanism to produce flares in active stars with close-in planets is described in this Section. It considers the transition from a non-linear force-free field to a potential magnetic field ($\alpha=0$) with the same boundary conditions at the photosphere, its energy $E_{\rm P}$ being the absolute minimum energy for the given boundary conditions. }
Given that the potential field has zero relative helicity by definition and that the helicity is conserved during the field evolution, the potential state is not accessible to the coronal field if its relative helicity is non-zero. {  In other words, the minimum energy $E_{\rm P}$ can be reached only if the non-linear field can get rid of all of its relative helicity. Since helicity cannot be dissipated during the field evolution, the only way to eliminate it is  by pushing it to the infinity by opening up all the field lines as discussed by, e.g., \citet{Flyeretal04}. The process that they envisage is based on the emergence of new magnetic flux from the stellar convection zone that steadily increases the magnetic helicity of the stellar corona until a threshold depending on the boundary conditions is reached beyond which no stable force-free equilibrium exists.} At that point, the field erupts producing a major flare with an associated coronal mass ejection that carries away the excess helicity \citep[cf.][]{Zhangetal06,ZhangFlyer08}. Since the orbital motion of the planet produces a modulation of the magnetic helicity of the stellar corona \citep{Lanza12}, the eruption can be triggered by the planet itself, if the magnetic helicity of the field gets sufficiently close to the threshold value. {  Note, however, that the process can operate also in stars without any close-in planet with the flare occurring when the helicity threshold is reached in the course of the field evolution produced by the emergence of new magnetic flux. The difference in the case of the stars with a close-in planet is the additional triggering mechanism associated with the planet that can operate when it is closer to the star, that is at the periastron of an eccentric orbit as observed in the case of the very active binary V773~Tau.} 

Considering for simplicity the case of an axisymmetric force-free coronal field, the accumulation of magnetic helicity leads to the formation of an azimuthal flux rope in the field configuration that increases its magnetic energy up to the point that  all the magnetic field lines can be opened.  The minimum energy of a field whose  lines have one end on the photosphere and the other at the infinity, that is the minimum energy of an open field with the same boundary conditions of the initial field, is called the Aly energy $E_{\rm A}$ of the field \citep[see][ and references therein]{Aly91,Sturrock91,LowSmith93,Flyeretal04}. In other words, the accumulation of helicity leads the field to reach an energy equal to $E_{\rm A}$ and at that point it can spontaneously erupt and get rid of all of its helicity. Now, the field can relax to the potential minimum energy state and release the maximum amount of magnetic energy, that is $\Delta E_{\rm max} = E_{\rm A} -E_{\rm P}$. For the non-linear force-free fields considered by \citet{Flyeretal04} and \citet{Zhangetal06}, the photospheric boundary conditions are those of a potential dipole field and $E_{\rm A} = 1.6616 E_{\rm P}$. {  To account for  different boundary conditions,  we now consider a specific family of force-free fields that allow us an analytic formulation of the problem and calculate their Aly energy.  In real cases, only some part of the coronal field lines may open and relax to the potential configuration. We considered the case of a complete opening of all the field lines because we want to estimate the maximum available energy. }

In spite of the simplicity of the defining equation (\ref{fff_def}), force-free fields are very complex mathematical objects and some simplifying assumptions are required for their analytical treatment. 
We shall consider the non-linear axisymmetric force-free model by \citet{LowLou90} in the implementation given by \citet{Wolfson95} that was already applied by \citet{Lanza12} to investigate star-planet interactions. {  We shall henceforth refer to the latter as the Wolfson field, while we shall use the name Low \& Lou fields to refer to the more general family of field configurations as introduced by \citet{LowLou90}.}

Assuming a reference frame with the origin at the barycentre of the star $O$ and the polar axis $\hat{z}$ along its rotation axis, the components of the magnetic field in a spherical polar coordinate system $(r, \theta, \phi)$ are: 
\begin{equation}
\vec B = \frac{B_{0} R^{2}}{r \sin \theta} \left[  \frac{1}{r} \frac{\partial A}{\partial \theta} \hat{\vec r} - \frac{\partial A}{\partial r} \hat{\vec \theta} + \frac{1}{R} Q(A)  \hat{\vec \phi} \right],
\label{lowloufield}
\end{equation}
where $B_{0}$ sets the intensity of the field at the surface of the star, $A(r, \theta)$ is the flux function, and $Q=Q(A)$  a scalar function that has a different functional form according to the specifically considered family of fields. Note that both $A$ and $Q$  are non-dimensional in our definition. {  The projections of the magnetic field lines in the $r$-$\theta$ plane coincide with the lines of constant $A$ because from Eq.~({\ref{lowloufield}) it follows: ${\vec B} \cdot \nabla A = 0$. }} \citet{LowLou90} consider a separable flux function of the form:
\begin{equation}
A(r, \theta) = (r/R)^{-n} f(x),
\label{wolfsondecay}
\end{equation}
where $x \equiv \cos \theta$, $n$ is a positive constant, not necessarily an integer, and $f$ is given by the differential equation:
\begin{equation}
(1 -x^{2}) f^{\prime \prime}(x) + n (n+1) f(x) + \lambda^{2} [f(x)]^{1+ 2/n} = 0,
\label{wolfson_bvp}
\end{equation}
that is solved in $[-1, 1]$ subject to the boundary conditions $f(-1)=f(1)=0$ with $\lambda^{2}$ as an eigenvalue.  {  All the field lines of the Low \& Lou fields are connected to the photosphere because if there were a closed field line detached from the photosphere, the continuous flux function $A(r, \theta)$ would have a local extremum at least in one point internal to that line, i.e., there will be a point where $\partial A / \partial r = \partial A / \partial \theta = 0$. This is not allowed given the mathematical form of $A$ as specified in Eq.~(\ref{wolfsondecay}). Therefore, Low \& Lou fields have no azimuthal flux rope and their energy  is always lower than that of the corresponding Aly field, whatever their boundary conditions, according to a theorem by \citet{Aly91} (see Sect.~\ref{results} for some numerical examples).} 
The scalar function $Q$  is given by:
\begin{equation}
Q(A) = \lambda A^{1 +1/n}, 
\end{equation}
while the force-free parameter $\alpha$ is:
\begin{equation}
\alpha= \frac{1}{R} \frac{dQ}{dA} = \frac{\lambda}{R} \frac{n+1}{n} \left( \frac{r}{R} \right)^{-1} [f(x)]^{{1}/{n}}. 
\label{alpha-def}
\end{equation}
We shall consider fields with $0 <n < 1$ because they have the slowest decay  with the distance (cf. Eq.~\ref{wolfsondecay}) and look for solutions of the boundary value problem for $f(x)$ 
that satisfy the conditions $f(-x) = f(x)$ in $[-1, 1]$ and $f^{\prime}(0) = 0$, following the method described by \citet{Wolfson95}. {  Note that the solution for $n=1$ corresponds to the potential dipole field, while that for $n=2$ to a quadrupole potential field. Solutions with $ 1 < n < 2 $ are generally characterized by a quadrupole topology, unless the azimuthal shear of the magnetic footpoints between the pole and the equator exceeds $\sim 90^{\circ}$ \citep{Wolfsonetal96}. Therefore, solutions with $0 < n < 1 $ are more suitable to represent the dipole-like configuration of the large-scale solar and stellar  coronae. 
 In the case of solar-type active stars, large-scale axisymmetric dipole-like configurations have often been revealed by Zeeman Doppler Imaging  \citep[][]{DonatiLandstreet09,Seeetal16}, thus supporting our choice of  $0 < n < 1$. By decreasing the parameter $n$ toward zero, it is possible to reproduce an increasing azimuthal shear of the footpoints of the field lines at different latitudes as expected from surface differential rotation, accompanied by a bulging out of the field  and a progressive localization of the electric current density close to the equatorial plane \citep[cf.][]{Wolfson95,Wolfsonetal96}. This allows us to model large-scale fields with a sizeable toroidal component at the photosphere as indeed observed by \citet{Petitetal08} in stars rotating faster than $\sim 12$~days. } 

The components {  of the Wolfson field} are:
\begin{eqnarray}
B_{r} & = &  -B_{0} (r/R)^{-(n+2)} f^{\prime} (x), \nonumber \\
B_{\theta} & = & n B_{0}  (r/R)^{-(n+2)} \frac{f(x)}{\sin \theta}, \label{wolfson_field} \\
B_{\phi} & = & \lambda B_{0}  (r/R)^{-(n+2)} \frac{[f(x)]^{1+ 1/n}}{\sin \theta}. \nonumber
\end{eqnarray}
{  We see that  fields with $ 0 < n< 1$, i.e.,  topologically equivalent to a dipole field, have a slower decay with distance from the star, while fields  topologically equivalent to higher order multipoles decrease faster with distance  making them much less relevant at the typical star-planet separations. }

The energy of any force-free field can be computed by means of  Eq.~(79) of \S~40 of \citet{Chandrasekhar61} that can be written for the space $V$ outside a spherical star as: 
\begin{equation}
\int_{V} \frac{B^{2}}{2{\mu}}\, dV =  \frac{1}{2{\mu}} R \int_{S(V)} (B_{r}^{2} - B_{\theta}^{2} - B_{\phi}^{2})\, dS, 
\label{chandra}
\end{equation}
where $S(V)$ is the surface of the star. By substituting Eqs.~(\ref{wolfson_field}) into Eq.~(\ref{chandra}) and considering that the field is independent of the azimuthal coordinate $\phi$, we find its magnetic energy: 
\begin{eqnarray}
E_{\rm M}  & = &\frac{1}{2{\mu}} \int_{V} B^{2}\, dV  =  \nonumber \\
\lefteqn{=\, \frac{\pi}{{\mu}} B_{0}^{2} R^{3}\!\! \int_{-1}^{1}\! \left\{ [f^{\prime}(x)]^{2} - n^{2} \frac{[f(x)]^{2}}{1-x^{2}} - \lambda^{2} \frac{[f(x)]^{2+2/n}}{1-x^{2}} \right\} dx.} 
\label{nonlinear_energy}
\end{eqnarray}
The minimum energy of the magnetic field in  $V$ is that of the potential field with the same radial component over the surface of the star $B_{r} = -B_{0} f^{\prime} (x)$. By applying the standard method to solve the Laplace equation with a prescribed normal component at the boundary of the domain $V$, we find:
\begin{equation}
E_{\rm P}  =  \frac{\pi}{{\mu}} B_{0}^{2} R^{3} \sum_{l=1}^{\infty} \frac{2l+1}{2(l+1)} \left[ \int_{-1}^{1} f^{\prime} (x) P_{l}(x)\, dx\right]^{2}\!\!,
\label{potential_energy}
\end{equation} 
where $P_{l}(x)$ is the Legendre polynomial of order $l$. 

Given its importance in our model, we compute the Aly energy for the field with the photospheric boundary conditions of the Wolfson field. Following the method in Appendix A of \citet{LowSmith93}, we find:
\begin{equation}
E_{\rm A} = \frac{\pi}{{\mu}} B_{0}^{2} R^{3} \left\{\int_{0}^{\pi} \! \! [f^{\prime}(\cos \theta)]^{2} \sin \theta \, d\theta - \!  \sum_{l=1}^{\infty} \frac{(4l+1) l}{2l+1} {\cal J}_{l}^{2} \right\},
\label{aly_energy}
\end{equation} 
where
\begin{equation}
{\cal J}_{l} \equiv \int_{0}^{\pi} A^{*}(R, \theta)\, P^{1}_{2l}(\cos \theta)\, d\theta,
\label{aly_energy_integrals}
\end{equation}
$P_{2l}^{1} (\cos \theta) = - dP_{2l} (\cos\theta)/d\theta $ being the associated Legendre functions of order $2l$, and the flux function $A^{*}(R, \theta)$ at the surface of the star is defined as \citep[cf.][App.~A]{LowSmith93}:
\begin{equation}
A^{*}(R, \theta) = \left\{  
\begin{array}{ll}
1+x-f(x), & \mbox{if $x \equiv \cos \theta \leq 0$,} \\
f(x)+x-1, & \mbox{if $x  >0$.} 
\end{array}
 \right.
\end{equation}
In the case of the Wolfson field, the energy is always smaller than the Aly limit and the spontaneous opening of all the field lines is not possible \citep[cf.][]{Wolfson95}. Nevertheless, we shall consider the possibility of a transition from the Aly state to the potential configuration for the Wolfson field  to illustrate how the maximum energy {  available to produce a flare can exceed  the limit of $\Delta E_{\rm max} = 0.662 E_{\rm P}$ found by \citet{Flyeretal04}}. This happens because the photospheric boundary conditions of the Wolfson fields are different from those of a potential dipole as assumed by \citet{Flyeretal04}. Of course, the Wolfson fields need some additional source of energy to reach the Aly state. It could be provided, for example, by the gravitational energy stored in a heavy plasma condensation in the corona \citep[e.g.,][]{LowSmith93,Lanza09}, but this implies to go beyond our force-free approximation, so we shall not investigate those  additional sources of energy. 

In the process described in this Section, we assume that  the  helicity modulation induced by the planet that triggers the field eruption takes place on a typical timescale $\tau_{\rm p} \sim L/v_{\rm rel}$, where $L$ is the typical lengthscale of the magnetic field and $v_{\rm rel}$ the relative velocity of the planet to the field lines. We assume that $\tau_{\rm p}$  is significantly longer than the Alfven transit time $\tau_{\rm A} \sim L/v_{\rm A}$,  so that our magnetohydrostatic models can still be applied. 
Once the instability is triggered, the time scale for the energy release is comparable to the Alfven transit time across the stellar corona because ideal MHD instabilities are generally invoked to account for flares. This is because of their much shorter development timescales in comparison to resistive instabilities in the high-conductivity environment of the stellar coronae \citep{Browningetal08,Lanza12}. Typical values of $\tau_{\rm A}$ range between $10^{2}$ and $10^{4}$~s in the coronae of stars with close-by planets \citep[cf. Sect.~3.1 in][and Sect.~\ref{applications} below]{Lanza12}.

\subsubsection{Magnetic energy available when the field is confined into a finite volume}
\label{energy_star_field}

In addition to the mechanism introduced in Sect.~\ref{star_field}, we consider another process that can lead to the release of magnetic energy, although by a smaller amount. It occurs when 
the field cannot get rid of its helicity {  opening up  its field lines.  
From a physical point of view, the confinement can be achieved for example  through the weight of an overlying atmosphere \citep[cf. Appendix~A in][]{ZhangLow03}. 
Configurations with closed field lines have been considered to explain the reduced angular momentum loss rate of stars hosting hot Jupiters \citep{Lanza10,Maxtedetal15} as well as the stability of plasma condensations in the coronae of stars hosting transiting planets \citep{Lanza09,Lanza14}\footnote{The latter were invoked to explain the observed correlation between planet surface gravity and stellar chromospheric fluxes  \citep{Figueiraetal14,Fossatietal15,Fossatietal17} because gravity rules the evaporation rate of the planets and thus the amount of material  available to form the condensations that absorb stellar flux in the chromospheric resonance lines.}. Numerical MHD simulations have confirmed that the interaction of the coronal field with a close-by planet tends to produce closed magnetic configurations \citep[e.g.,][]{Cohenetal09,Cohenetal10}.} 
 In this case, a non-linear field can  reach a minimum energy configuration if it is confined between the photosphere and some limit radius $r_{\rm L}$ with its radial component vanishing over the sphere $r=r_{\rm L}$, i.e., $B_{r} (r_{\rm L}, \theta, \phi) = 0$. For this system, the minimum energy configuration is a linear force-free field with the same relative helicity of the initial non-linear field  {  \citep[cf.][]{Dixonetal89,WiegelmannSakurai12}.} 

{  The transition from a non-linear to a linear force-free field in a confined domain was considered by \citet{ZhangLow03} to estimate the free magnetic energy made available by the emergence of new magnetic flux in a previous solar active region or by \citet{RegnierPriest07} to compute the energy available to power solar flares,  in both cases finding a good agreement with the observations. } Note that a linear force-free field extending to the infinity would have an infinite energy \citep{ChandrasekharKendall57}. Therefore, we need to consider a linear field within a confined domain. 
We shall follow an approach similar to the above mentioned works and assume that the stellar coronal field is confined within a sphere of radius $r_{\rm L} $. In the present Section, we shall illustrate how to compute the energy of the linear field with the same boundary conditions and relative helicity of a generic non-linear force-free field. The numerical results will be presented in Sect.~\ref{results} in the case of a specific non-linear field for different values of the radius of the outer  boundary $r=r_{\rm L}$. We shall find that the energy of the linear field with the same relative helicity becomes smaller than the energy of the non-linear field only if $r_{\rm L}$ is larger than a certain value $r_{\rm E}$ that depends on the specific configuration of the non-linear field. In other words, only if $r_{\rm L} \geq r_{\rm E}$ the transition from the non-linear to the linear field will release energy and can occur spontaneously. 

The case when $r_{\rm E}$ is smaller than the orbital separation of the planet at the periastron $a (1-e)$, where $a$ is the semimajor axis and $e$ the eccentricity of the orbit, is particularly relevant because it gives a final linear field configuration leading to a minimum energy dissipation rate at the boundary of the planetary magnetosphere. Specifically, the energy released by  reconnection between the stellar coronal field and the magnetic field of the planet at the boundary of its magnetosphere is:
\begin{equation}
\frac{dE_{\rm M}}{ dt} \propto \frac{B_{\rm m}^{2}}{2{\mu}} A_{\rm int} v_{\rm rel},
\end{equation}
where $B_{\rm m}$ is the field strength at the boundary of the magnetosphere, $A_{\rm int}$ the area of interaction comparable with the cross-section of the magnetosphere, and $v_{\rm rel}$ the relative velocity between the coronal magnetic field lines and those of the planet magnetosphere \citep[cf. Sect.~4.1 of][]{Lanza09}. Neglecting the radius of the planetary magnetosphere in comparison to the orbital separation, $dE_{\rm M}/dt$ is minimized when the stellar field is confined within a sphere of radius $r_{\rm L} \leq a(1-e)$. In this case, the coronal field lines simply slide over  the planetary field lines without any velocity component that pushes them toward each other as in the case of a coronal field extending beyond the orbital distance. In other words, this  closed configuration minimizes the amount of magnetic energy transported into the reconnection region per unit time, thus giving the lowest dissipation rate.

The coronal field is not static and we can assume that it makes transitions from the linear (and closed) to the non-linear (and partially open) force-free configurations and viceversa because of the continuous pumping of relative helicity by the emergence of new magnetic flux through the photosphere and the reconnection with the planetary field lines.  The spontaneous formation of current sheets inside the non-linear configurations \citep[e.g.,][]{Parker94,PontinHuang12} or the perturbations by the planet can  trigger a transition to a linear state, thus promoting a global energy release that can power a stellar flare. For example, the planet can perturb a configuration close to the threshold for the development of the kink instability as assumed by a flare model proposed by \citet{TorokKliem05}. This triggering process is not necessarily distinct from the one discussed in Sect.~\ref{star_field}  because a kink-unstable configuration can be produced by the accumulation of magnetic helicity in the stellar corona that leads to an increase of the twist of the field lines. 

We shall now apply our model to compute the energy available in the non-linear-to-linear field transition, {  that is when the field cannot open its lines and get rid of its helicity. }

To find the linear force-free field with the minimum energy corresponding to an initial non-linear force-free field, we shall consider the linear field with the same radial component at the stellar surface $r=R$, confined by the magnetic surface at $r=r_{\rm L}$, and with the same relative magnetic helicity of the non-linear field. These constraints are sufficient to define uniquely the minimum energy field with its constant force-free parameter $\alpha$. 

We start with the non-linear field and compute its relative magnetic helicity $H_{\rm R}$. For simplicity, we specialize our model to the case of a non-linear axisymmetric field. 
In this case, we can apply a formula found by \citet{Berger85} and \citet{Prasadetal14}, where the  domain $V$ is the space outside the stellar surface, i.e., the sphere of radius $r=R$: 
\begin{equation}
H_{\rm R} = 2 \int_{V} A_{\phi} B_{\phi}\, dV,
\label{rel_hel}
\end{equation}
where $\vec A$ is the vector potential of the non-linear field $\vec B$, i.e., ${\vec B} = \nabla \times {\vec A}$. Considering the specific fields in Sect.~\ref{star_field}, by comparing this equation with Eq.~(\ref{lowloufield}) and noting that no quantity depends on $\phi$, we find: 
\begin{equation}
A_{\phi} = \frac{B_{0} R^{2} A(r, \theta)}{r \sin \theta}, 
\end{equation}
where $A(r, \theta)$ is the flux function introduced in Eqs.~(\ref{lowloufield}) and~(\ref{wolfsondecay}). 
By performing the integration over the volume $V$ outside the star, we find the relative helicity of the non-linear force-free field of Wolfson as: 
\begin{equation}
H_{\rm R(NLFF)} = \frac{2 \pi B_{0}^{2} R^{4} \lambda}{n} \int_{-1}^{1} \frac{[f(x)]^{2 +1/n}}{1-x^{2}}\, dx, 
\label{nonlinear_helicity}
\end{equation}
that is  the dimensional version of the analogous formula in \citet{Prasadetal14}. 

In the second step, we consider the generic axisymmetric linear force-free field as given by \citet{ChandrasekharKendall57}, i.e.:
\begin{equation}
{\vec B} = \sum_{l=0}^{\infty} {\vec B}_{l},
\end{equation}
 with: 
\begin{equation}
{\vec B}_{l} = {\vec S}_{l} + {\vec T}_{l}\, , 
\end{equation}
where ${\vec S}_{l}$ and ${\vec T}_{l}$ are the poloidal and toroidal components of the field that are orthogonal to each other for a given order $l$ as well as orthogonal for different orders $l$ and $l^{\prime}$ \citep[cf.][App.~III, \S~129]{Chandrasekhar61}; moreover, ${\vec S}_{l} = \alpha^{-1} \nabla \times {\vec T}_{l}$ with $\alpha$ constant and independent of $l$. In the case of an axisymmetric field, their expressions are:
\begin{eqnarray}
{\vec S}_{l} & = & \frac{l(l+1)}{\alpha r} Z_{l} (\alpha r) P_{l} (x) \hat{\vec r} - \frac{1}{\alpha r} \frac{d}{dr} [r Z_{l} (\alpha r)] P_{l}^{1} (x) \hat{\vec \theta} \nonumber \\
{\vec T}_{l} & = & Z_{l} (\alpha r) P_{l}^{1}(x) \hat{\vec \phi},
\label{poloidal_toroidal_comps}
\end{eqnarray}
where the functions $Z_{l}(\alpha r)$ are given by:
\begin{equation}
Z_{l}(\alpha r) = \left( \frac{\pi}{2\alpha r} \right)^{1/2} \left[ c_{l} J_{l+1/2} (\alpha r) + d_{l} J_{-(l+1/2)}(\alpha r) \right];
\label{z_function}
\end{equation}
here $c_{l}$ and $d_{l}$ are coefficients depending on the boundary conditions and $J_{k}$ is a Bessel function of the first kind of order $k$. 

We can evaluate the total energy of the field in the volume $V^{\prime}$ between the concentric spheres $r=R$ and $r=r_{\rm L}$  by applying the method  in \citet{ChandrasekharKendall57} and considering that $B_{\rm r} = B_{\phi} = 0$ for $r=r_{\rm L}$. For a generic order $l$, we find: 
\begin{equation}
\int_{V^{\prime}} S_{l}^{2}\,dV^{\prime} = \int_{V^{\prime}} T_{l}^{2}\, dV^{\prime} + \frac{1}{\alpha} \int_{S} {\vec S}_{l} \times {\vec T}_{l} \cdot \hat{\vec r}\, dS, 
\label{energy_sl}
\end{equation}
where $S$ is the sphere $r=R$, i.e., the surface of the star. By making use of this equation, summing over all the orders $l$, and taking into account the orthogonality properties of the poloidal and toroidal fields, we find the expression for the total magnetic energy of the linear force-free field in the volume $V^{\prime}$:
\begin{equation}
E_{\rm LFF} = \frac{1}{2{\mu}} \sum_{l=1}^{\infty} \left(  \int_{V^{\prime}} 2T_{l}^{2}\, dV^{\prime} + \epsilon_{l} \right),
\label{ener_lff0}
\end{equation}
where 
\begin{equation}
\epsilon_{l} \equiv \frac{2l+1}{l(l+1)} \pi R^{3} \int_{-1}^{1} B_{lr} P_{l}\, dx \int_{-1}^{1} B_{l\theta} P_{l}^{1}\, dx
\label{epsilon_coeff}
\end{equation}
is the surface contribution to the energy of the order $l$ that appears in Eq.~(\ref{energy_sl}), $B_{rl}$ and $B_{l\theta}$ being the radial and meridional components of the poloidal field ${\vec S}_{l}$ evaluated at $r=R$, respectively. 

The relative helicity of the linear force-free field can be computed by Eq.~(\ref{rel_hel}) choosing the gauge in which ${\vec A} = \alpha^{-1} {\vec B}$, that gives:
\begin{equation}
H_{\rm R(LFF)} = \frac{2}{\alpha} \int_{V^{\prime}} B_{\phi}^{2} \, dV^{\prime} = \frac{2}{\alpha} \sum_{l=0}^{\infty} \int_{V^{\prime}} T_{l}^{2} \, dV^{\prime}.
\label{helicity_linear}
\end{equation}
Substituting into Eq.~(\ref{ener_lff0}), we obtain:
\begin{equation}
E_{\rm LFF} = \frac{1}{2{\mu}} \left( \alpha H_{\rm R(LFF)} + \sum_{l=0}^{\infty} \epsilon_{l} \right), 
\label{linear_energy_eps}
\end{equation}
that allows us to express  the total energy of the linear force-free field in terms of its relative helicity and the surface contributions $\epsilon_{l}$. 

The  integral giving the energy of the toroidal field can be calculated by considering the orthogonality of the toroidal fields with different $l$  and those of the associated Legendre polynomials, that is:
\begin{equation}
\int_{V^{\prime}} B_{\phi}^{2}\, dV^{\prime} = 4 \pi \sum_{l=0}^{\infty}  \frac{l (l+1)}{2l+1} \int_{R}^{r_{\rm L}} r^{2} [Z_{l} (\alpha r)]^{2}\, dr.  
\label{toroidal_energy}
\end{equation}

The formulae given above allow us to find the linear force-free field with the same relative helicity of an axisymmetric non-linear field, bounded by the sphere $r=r_{\rm L}$, and with the same radial component at the surface of the star. The latter boundary condition is required by the continuity of the magnetic flux across the surface of the star. It poses a constraint on the value of $Z_{l}(\alpha R)$ that together with the vanishing of the radial field at $r_{\rm L}$, i.e., $Z(\alpha r_{\rm L}) = 0$, can be used to find the coefficients $c_{l}$ and $d_{l}$ to specify the radial functions $Z_{l}$, if $\alpha$ is given. Specifically, $c_{l}$ and $d_{l}$ are found by solving the linear system:
\begin{equation}
\left\{
\begin{array}{l}
c_{l} J_{l+1/2} (\alpha R) + d_{l} J_{-(l+1/2)} (\alpha R)  =  p_{l} \\ 
c_{l} J_{l+1/2} (\alpha r_{\rm L}) + d_{l} J_{-(l+1/2)} (\alpha r_{\rm L})   =  0, 
\end{array}
\right.
\label{linear_field_coeffs}
\end{equation}
where 
\begin{equation}
p_{l} = \sqrt{\frac{2}{\pi}} \frac{2l+1}{2l(l+1)} (\alpha R)^{3/2}  \int_{-1}^{1} B_{r}(R,x) P_{l}(x) \, dx,
\label{pl_coeff}
\end{equation}
and $B_{r}(R, x)$ is the radial component of the non-linear force-free field at the surface. In the case of the Wolfson field, $B_{r} (R,x) = -B_{0} f^{\prime}(x)$ (cf. Eq.~\ref{wolfson_field}).  

To find the linear field with the same helicity $H_{\rm R}$ of the non-linear field, we iterate on the value of $\alpha$ until the condition $ H_{\rm R(LFF)} = H_{\rm R(NLFF)}$ is verified, using Eq.~(\ref{helicity_linear}) to compute $H_{\rm R(LFF)}$ for a given $\alpha$.  Actually, there are infinite values of $\alpha$ that satisfy the condition $H_{\rm R(LFF)} = H_{\rm R(NLFF)}$, as illustrated by \citet{Berger85}. This happens because both the energy and the relative helicity of the linear field diverge for the values $\alpha=\alpha_{\rm e}$ that make $Z(\alpha_{\rm e} R)$ = 0. They can be considered as  the eigenvalues of our problem. The plots of the energy and relative helicity versus $\alpha$ show an infinite number of branches along which $H_{\rm R(LFF)}$ takes all the values between a minimum and infinity \citep[cf. Fig.~1 in][and Fig.~\ref{Fig3} in Sect.~\ref{results}]{Berger85}. We always consider the first of those branches that starts from zero  relative helicity because the other branches have increasingly larger minima of $H_{\rm R(LFF)}$ that makes sometimes impossible to find a solution to our problem.  

{  In addition to the Wolfson fields} or the Low \& Lou fields, other non-linear force-free field models have been introduced in the literature \citep[e.g.][]{TitovDemoulin99,Flyeretal04}. The axisymmetric fields of \citet{Flyeretal04} are particularly suitable to describe the large-scale magnetic configuration of the stellar field at the distance of close-by planets because they do not decay rapidly with distance as in the case of localized fields {  and can allow for the presence of a flux rope. However, they assume a dipolar photospheric boundary condition and can be treated only numerically, so we limit ourselves to the case of Wolfson fields that allow for more general boundary conditions and can be treated in a fully analytic way. Moreover, }
Low \& Lou fields provide an approximate asymptotic representation of the fields of Flyer et al. when they have no azimuthal magnetic flux rope \citep{Flyeretal04,Lanza12}, given that their $\partial A(r, \theta)/\partial r$ has a constant negative sign, whereas the partial derivatives of $A(r, \theta)$ vanish on the axis of a  flux rope (cf. Sect.~\ref{star_field}).  
On the other hand, when an azimuthal rope of magnetic flux is present, the energy of the field is generally greater than in the case without a flux rope for the same boundary conditions \citep[cf. Fig.~7 and Fig.~8 in][]{Flyeretal04}. Therefore, our simple model provides a lower limit for the energy that can be released in the transition between the non-linear Flyer et al.'s dipole-like fields and the linear force-free fields, when the former do not have enough energy to open up all their field lines. 

{  For the mechanism introduced in this Section, we again assume that the timescale of the triggering associated with the planet $\tau_{\rm p}$ is much longer than the Alfven transit time $\tau_{\rm A}$ across the field configuration in order for our magnetostatic approximation to be valid. On the other hand, the timescale of the dissipation of the magnetic helicity by ideal MHD instabilities and reconnection processes is much longer than $\tau_{\rm A}$ \citep[cf. Sect.~2.4 in][]{Lanza12,Berger84}, thus conserving the relative helicity during the considered field transitions that lead to an energy release without opening up the field lines.}

\subsection{The case of weakly active stars} 
\label{potential_fields}

The models discussed so far are mainly suited for application  to very active stars with closed coronal field lines up to the distance of their close-by planets. In the case of moderately or weakly active stars, such as the majority of planet hosts,  we refer to the model by \citet{Seeetal17} who assume that the coronal field is potential and has closed field lines up to a radius $r = r_{\rm SS}$ that defines the so-called source-surface of the stellar wind \citep[cf.][]{AlthusserNewkirk69}. On that surface the potential is constant, so the outer field is purely radial and it is assumed to remain so also for $r > r_{\rm SS}$. 
  For the Sun, $r_{\rm SS} \sim 2.5\,R$, while for stars with spectropolarimetric detections of the surface fields, \citet{Seeetal17} adopt $r_{\rm SS} \sim 3.4\,R$ because their field strengths $B_{0}$ range between the solar value ($\approx 1-3$~G) and approximately ten times the solar value. Therefore, it is justified to assume that the field lines are open and radially directed at the typical orbital distances of close-by planets. Their deviation  from the radial direction induced by the rotation of the star, leading to the formation of the Parker's spiral of the interplanetary field, is negligible at the distance of close-by planets \citep{WeberDavis67}. Note that when the field is purely radial,  its strength varies as $B(r) \propto r^{-2}$. 
  
In the case of a potential field, the configuration is always in the state of absolute minimum energy for the given boundary conditions and has a zero relative helicity \citep{BergerField84}. Therefore, no energy can be released by the mechanisms considered in Sects.~\ref{star_field} and~\ref{energy_star_field} and a different process must be considered that we shall investigate in the next section. 
  
\subsubsection{The magnetic field of the planetary magnetosphere and its interaction with the stellar field}
\label{magnetospheric_effect}

We investigate the perturbation of the stellar potential field (cf. Sect.~\ref{potential_fields}) by the planetary magnetosphere to evaluate the energy variation when the distance of the planet from the star changes during its orbital motion or the planet moves through regions of different field intensity. In the present model, the stellar field can be non-axisymmetric \citep[cf.][]{AlthusserNewkirk69}, thus the planet can encounter a region of relatively strong field such as a coronal streamer during its motion through the outer stellar corona. 

Most of the parameters governing  planetary magnetospheres are presently unknown, notably we have no direct measurement of the planetary magnetic fields yet \citep[e.g.,][]{Vidottoetal10,Cauleyetal15,Rogers17}. Therefore, we prefer to use a very simplified model that assumes a prescribed geometry for the magnetopause, that is the surface separating the planetary magnetosphere from the stellar coronal or wind magnetic field. This has been done by, e.g., \citet{Griessmeieretal04}, who adopted a slightly modified version of the model proposed by \citet{Voigt81} for the Earth's magnetosphere; \citet{Khodachenkoetal12}, who assumed a paraboloid of revolution to describe the magnetopause; or \citet{Lanza12}, who adopted a spherical surface. 

We adopt a simplified version of the model by \citet{Voigt81} because it can be adapted for both sub-Alfvenic and super-Alfvenic flow regimes at the location of the exoplanets, while other models, for instance that of \citet{Khodachenkoetal12}, were designed only for a super-alfvenic regime. The regime of the stellar wind at the distance of the exoplanets is not directly measured, but extrapolations of the \citet{WeberDavis67} model and numerical simulations suggest that they are  in a sub-alfvenic regime in most of the cases, i.e., the velocity of the stellar wind $v_{\rm w}$ is smaller than the local Alfven velocity $v_{\rm A}$ owing to the close distance of the planets \citep{Sauretal13,Strugareketal15}. The regime observed in the Solar System is super-alfvenic, that is the wind velocity is faster than the local Alfven speed leading to the formation of a bow shock at the magnetopause in the case of a magnetized planet. In this case, the magnetic field lines of the stellar wind and of the planetary magnetosphere have no or little connection and the magnetosphere can be considered as an obstacle to the wind flow. 

On the other hand, in  the sub-alfvenic regime, there is no shock at the surface of separation and the magnetic field lines of the planet and  the wind  can partially reconnect, especially when the stellar field is potential \citep[cf.][]{Lanza13}. Perturbations excited along those lines can travel back to the star because their characteristic velocity is of the order of the Alfven velocity, thus faster than the wind. Indeed, most of the energy of the perturbation is channelled along the field lines giving rise to the so-called Alfven wings \citep[][]{Preusseetal06,Sauretal13,Strugareketal15}. The power transported by the Alfven wings has been computed by, e.g., \citet{Sauretal13}. The magnetic stresses produced by the orbital motion of the planet induce a steady dissipation inside any loop interconnecting the star and the planet as discussed by \citet{Lanza13}. Nevertheless, being interested in the processes that can store energy to be released in stellar flares, we shall not consider Alfven waves or energy dissipation due to magnetic stresses because these processes are usually steady   and not  impulsive  as the energy release associated with a flare. 

We focus on the effect of the planetary magnetosphere on the energy of the stellar outer field, neglecting the kinetic energy of the stellar wind because it is of the order of $(v_{\rm w}/v_{\rm A})^{2} \ll 1$ in the assumed sub-Alfvenic regime. {  For simplicity, we neglect also the variation of the energy of the planetary magnetosphere during the orbital motion of the planet, because the magnetospheric field is not completely connected to the stellar field (see below), thus its energy may not be available to power a stellar flare. We anticipate that the energy of the stellar coronal field is decreased by the presence of the planetary magnetosphere, the effect being larger when the planet comes closer to the star (see below Eq.~\ref{energy_var_magsp}). Similarly, the energy of the planetary magnetosphere decreases when the planet gets closer to the star, as we show in Appendix~\ref{app1}.  In other words, the total magnetic energy of the star-planet system decreases when the planet comes closer to the star, making it possible for the stellar coronal field to power a stellar flare.} 

Since the model by \citet{Voigt81} is  magnetostatic, we cannot include the effects of the Alfven waves in our model, but we can consider the reconnection between the stellar magnetic field and that of the planet. If we indicate with $V_{\rm e}$ the volume outside the star and the planetary magnetosphere of volume $V_{\rm m}$, that is $V = V_{\rm e} \cup V_{\rm m}$, the magnetic field ${\vec B}_{\rm e}$ in $V_{\rm e}$ can be written as \citep[cf. eq.~(3.23) of][]{Voigt81}:
\begin{equation}
{\vec B}_{\rm e} = {\vec B} + (1-C_{\rm imf}) {\vec B}_{\rm cfi} + C_{\rm d} {\vec B}_{\rm s}, 
\end{equation}
where ${\vec B}$ is the unperturbed stellar field, i.e.,  without the magnetosphere, ${\vec B}_{\rm cfi}$  the field produced by the Chapman-Ferraro currents that flow inside the infinitesimally thin magnetopause, ${\vec B}_{\rm s}$ the field inside the magnetosphere that includes the field of the planetary dynamo as well as that of the ionospheric currents and of the currents in the magnetospheric tail; $C_{\rm imf}$ and $C_{\rm d}$ are numerical coefficients between 0 and 1 that specify the efficiency of the reconnection between the stellar field and the planetary field across the magnetopause. In the simplified model by \citet{Voigt81}, both coefficients are assumed to be constant and are used to specify the boundary conditions at the magnetopause to compute the Chapman-Ferraro fields. For the Earth magnetosphere, a fit to the observations gives: $C_{\rm imf} \sim 0.9$ and $C_{\rm d} \sim 0.1$. Note that this small value of $C_{\rm d}$  was obtained in the super-alfvenic regime characteristic of the Earth's magnetosphere, while most of the close-by exoplanets are in a sub-alfvenic regime. Nevertheless, to simplify our treatment, we assume $C_{\rm d}=0$ that corresponds to assume that the flux of the internal field ${\vec B}_{\rm s}$ across the boundary of the magnetosphere is zero, thus avoiding problems associated with our ignorance of the planetary field and of the ionospheric and tail currents in the case of exoplanets. {  Note that our assumption $C_{\rm d}=0$ affects the field in the volume $V_{\rm e}$ outside the magnetopause, but it does not prevent the formation of the magnetopause itself with its Ferraro-Chapman currents,  provided that the magnetic field lines of the magnetospheric field $B_{\rm s}$ remain confined into the magnetopause. Conversely, by assuming $C_{\rm imf} \not=0$, we include the flux of the stellar field ${\vec B}$ across  the magnetopause and allow for its effects on the Chapman-Ferraro field outside the magnetopause. The parameter $C_{\rm imf}$, as well as $C_{\rm d}$ in a more general Voigt model, can vary because of the changing distance along an eccentric orbit, or because the planet encounters different regions of the stellar wind corresponding to sub- or super-alfvenic regimes, or due to the time variability of the wind itself \citep[cf.][]{Cohenetal14,Strugareketal15,Nicholsonetal16}.  Our formulae remain valid with the instantaneous value of $C_{\rm imf}$ provided that the timescale of the parameter change is much longer than the Alfven transit time across the magnetosphere. } 

The energy $E_{\rm e}$ of the stellar field in the volume $V_{\rm e}$ is given by:
\begin{eqnarray}
2{\mu} E_{\rm e}  = \int_{V_{\rm e}} \!\! {\vec B}_{\rm e}^{2} \, dV   =  \int_{V_{\rm e}} \!\! [{\vec B} - (1-C_{\rm imf}) \nabla u_{\rm cfi} ]^{2}  dV = \nonumber \\
 =  \int_{V_{\rm e}}\!\! B^{2} \,dV \! - \! (1-C_{\rm imf})\!\!\! \int_{V_{\rm e}}\!\!\! \nabla u_{\rm cfi}\cdot \left[ 2{\vec B} - (1- C_{\rm imf})\nabla u_{\rm cfi} \right] dV,  
 \label{energy_ve}
\end{eqnarray}
where $u_{\rm cfi}$ is the potential generated by the Chapman-Ferraro currents outside the magnetopause such that ${\vec B}_{\rm cfi} = -\nabla u_{\rm cfi}$. It satisfies the Laplace equation $\nabla^{2} u_{\rm cfi} = 0$ with closed boundary conditions on the surface $S_{\rm m}$ of the magnetopause, i.e.,  $\partial u_{\rm cfi}/\partial n = {\vec B} \cdot \hat{\vec n}$, where $\hat{\vec n}$ is the normal to $S_{\rm m}$ \citep[cf. Sect.~3 of][]{Voigt81}. Using identities for the divergence of a vector field, we find that: $\nabla \cdot \left\{ u_{\rm cfi} \left[ 2 {\vec B} - (1- C_{\rm imf}) \nabla u_{\rm cfi} \right] \right\} = \nabla u_{\rm cfi} \cdot \left[ 2 {\vec B} -(1-C_{\rm imf}) \nabla u_{\rm cfi} \right]$ because ${\vec B}$ is solenoidal and $u_{\rm cfi}$ satisfies the Laplace equation. Therefore, we can use the divergence theorem to rewrite Eq.~(\ref{energy_ve}) as: 
\begin{eqnarray}
2 {\mu} E_{\rm e} \! \!& = & \! \! \! \int_{V_{\rm e}}\!\!\! B^{2}\, dV + \nonumber \\ 
& &  -  \, (1- C_{\rm imf}) \!\! \int_{S(V_{\rm e})} \!\!\! \!\! u_{\rm cfi} \left[ 2 {\vec B} - (1- C_{\rm imf}) \nabla u_{\rm cfi} \right] \cdot \hat{\vec n}\, dS = \nonumber \\ 
 \lefteqn{= \int_{V_{\rm e}} \!\!\! B^{2}\, dV- (1- C_{\rm imf}^{2}) \! \int_{S_{\rm m}}\!\! u_{\rm cfi} {\vec B}\cdot \hat{\vec n} \, dS,}
 \label{energy_ve1}
\end{eqnarray}
where we made use of the boundary condition for $u_{\rm cfi}$ and considered its rapid decay with distance from the planet. This makes the contribution of the integral on the surface of the star negligible leaving only the integral on the surface $S_{\rm m}$ of the magnetosphere when we consider the integral over the whole boundary $S(V_{\rm e})$ of the volume $V_{\rm e}$. 

\begin{figure}
\flushleft{ 
 \includegraphics[width=9cm,height=8cm]{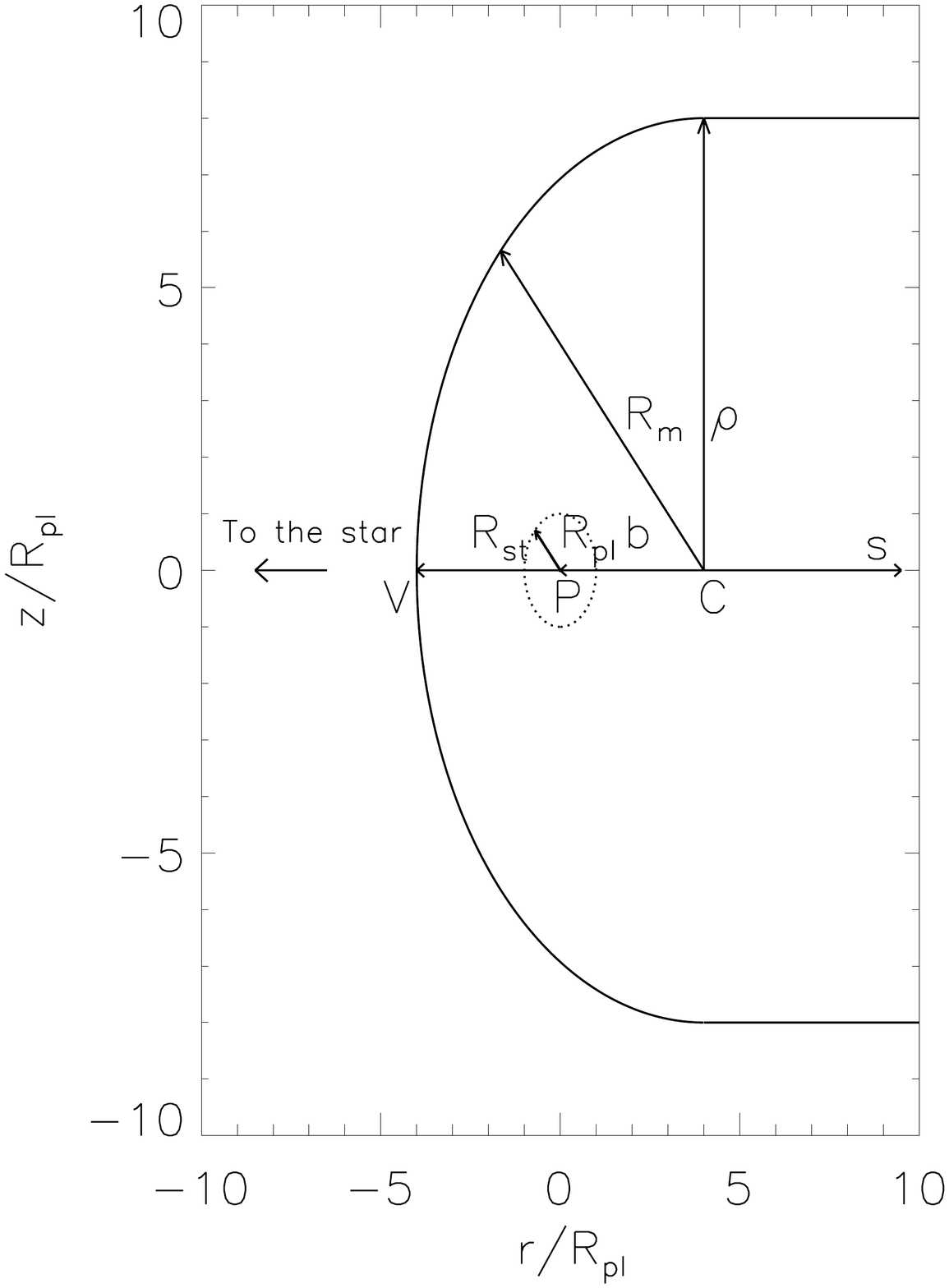}} 
   \caption{Cross-section of the magnetosphere in the model by \citet{Voigt81}. The radii of the hemispherical head and of the cylindrical tail are equal and are indicated with $R_{\rm m}$, while the radius of the planet is $R_{\rm pl}$ and its cross section is indicated by the dotted line. The cylindrical radius $\rho$ from the axis of the tail and the abscissa $s$ along the same axis are noted.  The centre of the planet $P$ is shifted towards the star by a distance $b$ with respect to the centre $C$ of the head of the magnetosphere. The stand-off distance $R_{\rm st}=R_{\rm m}- b$ is the length of the segment $PV$, where $V$ is the vertex of the magnetosphere pointing towards the star. The distance $r$ from the planet and the distance from the orbital plane $z$ are indicated on the plot box  in units of the planet radius. }
              \label{Fig1}%
\end{figure}

The magnetopause has a fixed geometry in our model consisting of a hemispherical head and a cylindrical tail with the same radius of the hemisphere $R_{\rm m}$ (see Fig.~\ref{Fig1}). \citet{Griessmeieretal04} assume that $R_{\rm m} = 2 R_{\rm st}$, where $R_{\rm st}$ is the standoff distance of the magnetopause from the centre of the planet that in our model is $R_{\rm st} = R_{\rm m} -b$, where $b$ is the separation between the centre of the planet and that of the head of the magnetosphere.  The stand-off distance is obtained by equating the magnetic pressure on both sides of the magnetopause, i.e. $B^{2} ({\vec r}_{\rm p})= B_{\rm s}^{2} (R_{\rm st})$, where  ${\vec r}_{\rm p}$ is the position vector of the planet. Following \citet{Griessmeieretal04}, we assume $B_{\rm s} (d) = 2 f_{0} B_{\rm p}(d)$, where $d$ is the distance from the barycentre of the planet, $f_{0}$ a shape factor, and $B_{\rm p} \propto d^{-3}$  the planetary magnetic field that we assume to decay as a dipolar field with the distance. The shape factor would be $f_{0} = 1.5$ for an ideal magnetosphere, but $f_{0} = 1.16$ gives a more realistic geometry and will be used here \citep{Griessmeieretal04}. By applying this simple model, we obtain:
\begin{equation}
R_{\rm m} = 2 \,(2f_{0})^{1/3} \left[ \frac{B_{\rm pl}}{B({\vec r}_{\rm p})}\right]^{1/3} \!\! R_{\rm pl},
\label{rm_eq}
\end{equation}
where $B_{\rm pl}$ is the magnetic field strength at the pole of the planet and $R_{\rm pl}$ its radius. 

By solving the Laplace equation with the boundary condition specified above, we obtain the  potential $u_{\rm cfi}$ and we can compute the energy with Eq.~(\ref{energy_ve1}). The expression of $u_{\rm cfi}$ is given in terms of the stellar field ${\vec B}$ by Eqs.~(5.15) and (5.30) of \citet{Voigt81} for the hemispherical head and the tail of the magnetosphere, respectively. They can be recast in the form:
\begin{equation}
u_{\rm cfi} = \left\{
\begin{array}{ll}
 \frac{1}{2} ({\vec B} \cdot \hat{\vec r}) \left( R_{\rm m}^{3}/r^{2} \right) & \mbox{for the hemispherical head,} \\ 
 ({\vec B} \cdot \hat{\vec \rho}) \left( R_{\rm m}^{2}/\rho \right) & \mbox{for the cylindrical tail},
\end{array}
\right. 
\label{ucfi}
\end{equation}
where ${\vec r}$ is the radius vector from the centre of the hemispherical head and ${\vec \rho}$ the cylindrical radius from the axis of the cylindrical tail of the magnetosphere. Note that $\hat{\vec r}$ and $\hat{\vec \rho}$ are opposite to the local unit normal $\hat{\vec n}$ on the hemispherical head and the cylindrical tail in Eq.~(\ref{energy_ve1}), respectively. 

By substituting Eqs.~(\ref{ucfi}) into Eq.~(\ref{energy_ve1}) and integrating over the head and the tail, we find: 
\begin{eqnarray}
2 {\mu} E_{\rm e} & = &  \int_{V_{\rm e}} \!\! B^{2}\, dV + \nonumber \\ 
\lefteqn{ -\, (1-C_{\rm imf}^{2}) \left[ \frac{\pi}{3} B^{2}({\vec r}_{\rm p}) R_{\rm m}^{3} + \pi R_{\rm m}^{2} \cos^{2} \xi \!\! \int_{r_{\rm p}}^{\infty}\!\!\!  B^{2}(s)\, ds \right],}
\label{energy_head_tail}
\end{eqnarray}
where $\xi$ is the angle between the direction of the North pole of the planet and that of the stellar field at the position of the barycentre of the planet ${\vec r}_{\rm p}$; note that $\xi = \pi/2$ when the field is directed from the star to the planet, while $\xi = -\pi/2$ when it is oppositely directed  \citep[cf.~Sect.~4][]{Voigt81}; the coordinate $s$ is the distance along the axis of the cylindrical magnetotail with $s=r_{\rm p}$ coinciding with the barycentre of the planet. In deriving Eq.~(\ref{energy_head_tail}), we assumed that the stellar field ${\vec B}$ is uniform over the head of the magnetosphere and each section of its tail orthogonal to the axis of the cylinder, while it varies along the tail axis $\hat{\vec s}$. However, the stellar field can be inhomogeneous over larger length scales, thus leading to an energy variation during the orbital motion of the planet. 

Specifically, we find an energy variation with respect to the unperturbed situation, i.e. that of a star without any planet, but with the same coronal field:
\begin{eqnarray}
2 {\mu} \Delta E & = &  -\pi \left\{ \left(1-\frac{1}{3}C_{\rm imf}^{2}  \right)  R_{\rm m}^{3} B^{2}({\vec r}_{\rm p})\, +  \right.  \nonumber \\ 
& & +  \left. \left[ \left( 1+(1-C_{\rm imf}^{2} \right) \cos^{2} \xi \right] R_{\rm m}^{2} \int_{r_{\rm p}}^{\infty} \!\! B^{2}(s)\, ds \right\}.
\label{delta_energy}
\end{eqnarray}
The variation of the strength of the stellar field with the distance from the star $r$ depends on the configuration of the field itself as we discussed in Sect.~\ref{star_field} and~\ref{potential_fields}. In the case of young and very active stars, we can use the Wolfson field model that gives: $B (r) = B_{0}  (r/R)^{-(n+2)}$ with $0 < n < 1$, where $B_{0}$ is the field at the surface of the star $r = R$; while for the field of old and weakly active stars beyond the source surface of their stellar wind: $B(r) = B(r_{\rm ss}) (r/r_{\rm ss})^{-2}$. {  Therefore, we may write: $B (r) = B(r_{\rm p}) (r/r_{\rm p})^{-(n+2)}$, where  $0 \leq n < 1$  allows for both of these different models. Note that the case $n=0$ in the Wolfson model corresponds to the radially directed field of a split monopole that is the same configuration as in the source-surface model of the wind field \citep[cf.][]{Wolfson95}. }

Evaluating the integral in Eq.~(\ref{delta_energy}), we finally find:
\begin{eqnarray}
2 {\mu} \Delta E  & = & -\pi R_{\rm m}^{3} B^{2}({\vec r}_{\rm p}) \left\{ \left(1-\frac{1}{3}C^{2}_{\rm imf} \right)  + \right. \nonumber \\ 
 & + & \left. \left[ 1+ (1-C_{\rm imf}^{2}) \cos^{2}\xi \right]  \frac{1}{2n+3} \left( \frac{r_{\rm p}}{R_{\rm m}}\right) \right\}. 
 \label{energy_var_magsp}
\end{eqnarray}
The energy variation is always negative, that is the planetary magnetosphere decreases the energy in comparison to the unperturbed stellar field. The variation is dominated by the second term in the outer braces because  $r_{\rm p}$ is significantly larger than $R_{\rm m}$. The maximum energy variation with respect to the star without any planet is obtained when $C_{\rm imf} = 0$ and $\cos \xi =0$, i.e., the magnetosphere is closed with the stellar field that does not penetrate across the magnetopause and the field is radially directed towards or away from the star  \citep[cf.][]{Voigt81}. For example, the stellar field is radial in the quadrupolar configuration considered by \citet{Strugareketal15} or when the  planet is inside a coronal streamer. On the other hand, when the stellar field has a dipolar configuration parallel or antiparallel to the planetary magnetic moment, $\cos^{2} \xi =1$, and the interconnection between the planetary and the stellar field parameterized by $C_{\rm imf}$ plays a relevant role. For given $\xi$ and $C_{\rm imf}$, considering the dependence of $R_{\rm m}$ on the stellar field $B({\vec r}_{\rm p})$ in Eq.~(\ref{rm_eq}) and the dependence of $B({\vec r}_{\rm p})$ on $r_{\rm p}$ given above, we find: $ \Delta E \propto (r_{\rm p}/R)^{-n-1}$ leading to the maximum energy variation at the periastron along an eccentric orbit.  

It is interesting to compare the energy variation in Eq.~(\ref{energy_var_magsp}) with that obtained by \citet{Lanza12} for a closed spherical magnetopause (cf. his Eq.~26), that is:
\begin{equation}
 {\mu} \Delta E_{\rm sp} = - \pi R_{\rm m}^{3} B^{2}({\vec r}_{\rm p}).  
\end{equation}
The ratio $\Delta E/ \Delta E_{\rm sp} \sim (r_{\rm p}/R_{\rm m})/(2n+3) \gg 1$ because $r_{\rm p} \gg R_{\rm m}$. The difference between the two models is due to the presence of the magnetospheric tail in that of \citet{Voigt81}. It leads to a remarkably larger energy variation in the stellar field because it has a larger volume than the spherical magnetosphere. 

The timescale $\tau_{\rm r}$ for the release of the energy in Eq.~(\ref{energy_var_magsp}) is equal to the time taken by the planet to cross the cross-section of the magnetosphere, i.e., $\tau_{\rm r} \sim 2R_{\rm m}/v_{\rm rel}$, {  where $v_{\rm rel}$ is the relative velocity between the planet and the magnetic field lines of the star. In the case of a prograde orbit, $v_{\rm rel} \simeq v_{\rm orb} -\Omega r_{\rm p}$, where $v_{\rm orb}$ is the orbital velocity of the planet and $\Omega$ the angular velocity of rotation of the star. If the stellar rotation period, $P_{\rm rot} = 2\pi/\Omega$, is significantly longer than the orbital period, $P_{\rm orb}$, then $v_{\rm rel} \simeq v_{\rm orb}$. }Considering the planet at the periastron, where $v_{\rm orb}$ is at its maximum, and a stellar field given by $B(r_{\rm p}) = B_{0}(r_{\rm p}/R)^{-2}$, we find:
\begin{equation}
\tau_{\rm r} \geq \frac{2}{\pi} P_{\rm orb} (2 f_{0})^{1/3}  \left( \frac{a}{R} \right)^{-1/3} \frac{(1-e)^{7/6}}{(1+e)^{1/2}} \left( \frac{B_{\rm pl}}{B_{0}} \right)^{1/3} \left( \frac{R_{\rm pl}}{R} \right). 
\label{release_time}
\end{equation}

\section{Results}
\label{results}

The theory introduced in Sect.~\ref{star_field} is applied to  the Wolfson field for three values of the index $n=0.1,0.25$, and $0.5$. The relative helicity $H_{\rm R(NLFF)}$ and energy $E_{\rm NLFF}$ of the non-linear field as well as the energy of the potential field $E_{\rm P}$ and of the Aly field $E_{\rm A}$ with the same boundary conditions are listed for the different values of $n$ in Table~\ref{table1}, respectively. The energy $E_{\rm NLFF}$ and  helicity $H_{\rm R(NLFF)}$  are obtained from  Eqs.~(\ref{nonlinear_energy}) and (\ref{nonlinear_helicity}), respectively, while the values of $E_{\rm P}$ and $E_{\rm A}$ come from Eqs.~(\ref{potential_energy}) and~(\ref{aly_energy})-(\ref{aly_energy_integrals}) by truncating the series at the order $l = 50$, respectively. 
\begin{table}
\caption{Relative helicity $H_{\rm R (NLFF)}$ and magnetic energy $E_{\rm NLFF}$ of the non-linear force-free field of \citet{Wolfson95} for different values of the parameter $n$ together with the energy $E_{\rm p}$ of the potential field and the energy $E_{\rm A}$ of the Aly field with the same photospheric boundary conditions.} 
\begin{center}
\begin{tabular}{rrccc}
\hline
$n$ & $H_{\rm R (NLFF)}$ & $E_{\rm NLFF}$ & $E_{\rm P}$ & $E_{\rm A}$ \\
 & $(B_{0}^{2} R^{4})$~ & $(B_{0}^{2} R^{3}/\mu)$ & $(B_{0}^{2} R^{3}/\mu)$ & $(B_{0}^{2} R^{3}/\mu)$ \\
 \hline
0.1 & 15.9648 & 5.9239 & 2.9217 & 6.3194 \\
0.25 & 12.2683 & 5.4719 & 3.2325 & 6.4139\\
0.5 & 8.2224 & 4.9031 & 3.6115 & 6.5928 \\ 
\hline
\end{tabular}
\end{center}
\label{table1}
\end{table}

The helicity of the Wolfson field is larger for smaller values of $n$ because the field has a greater shear. Its energy is always smaller than the Aly limit, thus the field cannot spontaneously open up its field lines, get rid of its helicity, and make a  transition to the minimum energy potential configuration with the same photospheric boundary conditions. The ratio $E_{\rm A}/E_{\rm P}$ ranges between $2.163$ and $1.825$ and decreases with increasing $n$. We remind that in the case of the fields of \citet{Flyeretal04}, having the  boundary conditions of a potential dipole, $E_{\rm A}/E_{\rm P} = 1.662$, thus demonstrating the crucial role played by the boundary conditions in establishing the value of this ratio. {  As in \citet{Flyeretal04}, it is interesting to consider the maximum upper bound $E_{\rm abs}$ for the energy of a force-free field with a prescribed $B_{r}$ at the photosphere, that is given by the first term in the r.h.s. of Eq.~(\ref{chandra}) \citep[cf. Eq. 25 in][]{Flyeretal04}:
\begin{equation}
E_{\rm abs} = \frac{1}{2\mu} R \int_{S(V)} B_{r}^{2} \, dS = \frac{\pi}{\mu} R^{3} B_{0}^{2} \int_{0}^{\pi} [f^{\prime}(\cos \theta)]^{2} \sin \theta \, d\theta,
\end{equation}
in the case of the Wolfson field. We see that it is equal to the first term appearing in the expression for the Aly energy, so that $E_{\rm A} < E_{\rm abs}$ (cf. Eq.~\ref{aly_energy}).  This upper bound is valid for any force-free field, including those with an azimuthal flux rope, and depends only on the boundary conditions at the surface of the star. In the case of the potential dipole boundary conditions of the fields of \citet{Flyeretal04}, $E_{\rm abs} = 2 E_{\rm P}$. In the case of fields with Wolfson's boundary conditions, i.e.,  $B_{r} = - B_{0} f^{\prime}(\cos \theta)$, we see that the ratio $E_{\rm abs}/E_{\rm P}$ ranges from $2.230$ to $2.036$ when $n$ increases from 0.1 to 0.5. Therefore, the excess of energy that can be stored into a force-free field with respect to the Aly limit is moderate also in the case of fields with Wolfson's boundary conditions, even in the eventuality they had a flux rope;  greater amounts of energy can be stored only in the presence of additional energy sources to confine the field. }

Next, we consider the transition of the Wolfson field to a linear force-free field confined within a radius $r_{\rm L}$ with the same relative helicity and photospheric boundary conditions  as discussed in Sect.~\ref{energy_star_field} and compute the energy made available in the transition.  

\begin{figure}
\flushleft{ 
 \includegraphics[width=9cm,height=6cm]{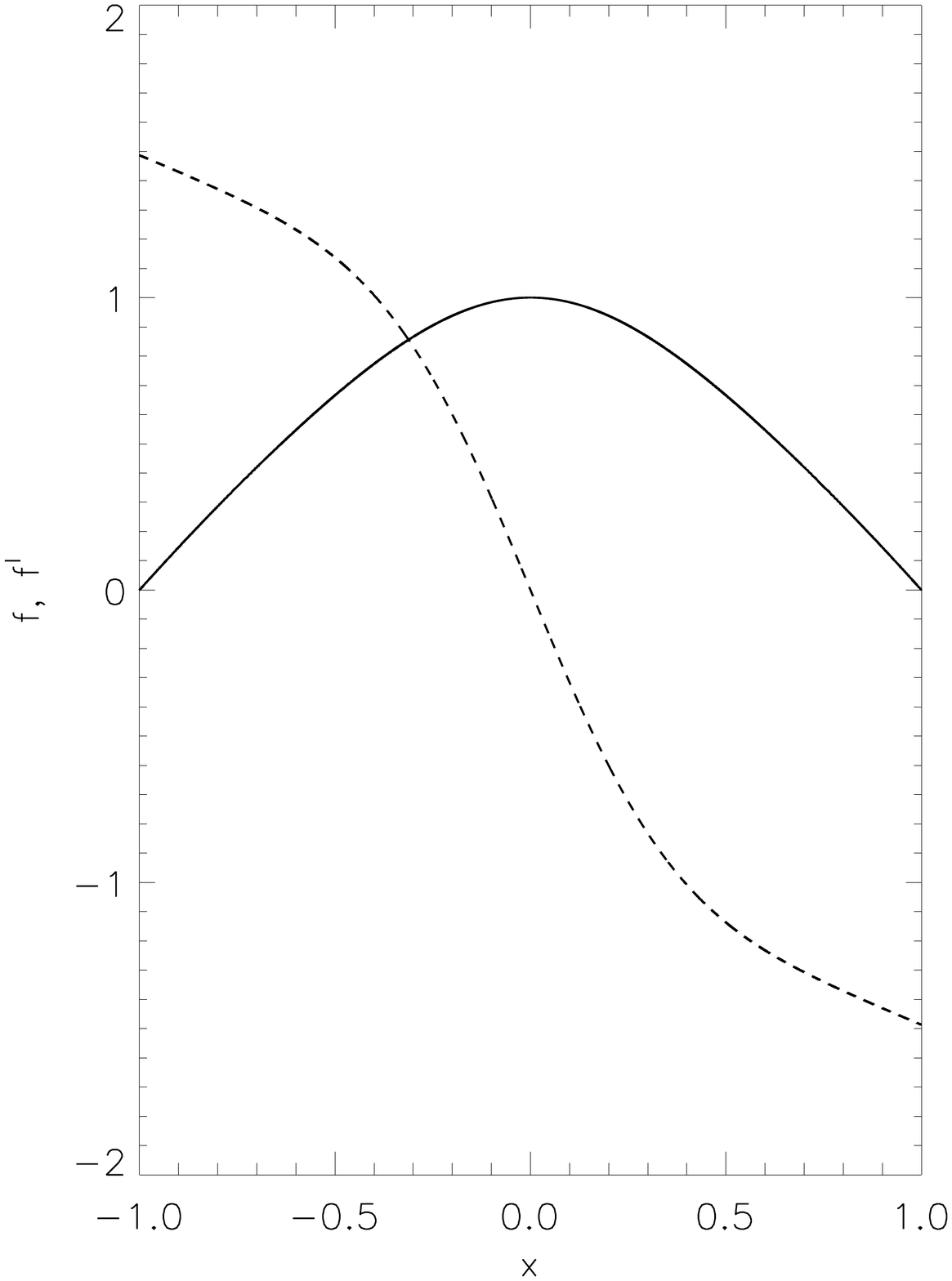}} 
   \caption{Function $f(x)$ (solid line) and its derivative $f^{\prime}(x)$ (dashed line) vs. $x = \cos \theta$ for the non-linear force-free field of \citet{Wolfson95} with $n=0.5$ as described in Sect.~\ref{star_field}.}
              \label{Fig2}%
\end{figure}

To provide an illustration of the method, we choose a non-linear Wolfson field with $n=0.5$. The function $f(x)$ and its derivative $f^{\prime}(x)$ with $x = \cos \theta$ are obtained by solving the boundary value problem as defined by Eq.~(\ref{wolfson_bvp}) with the boundary conditions $f(-1) = f(1) = 0$. The eigenvalue $\lambda = 0.90743$, while the function $f$ is plotted in Fig.~\ref{Fig2} together with its derivative. 

The radial component of the magnetic field at the photosphere is given by $B_{r} (R, x) = -B_{0} f^{\prime} (x)$ as follows from the first of the equations (\ref{wolfson_field}). The magnetic energy of the non-linear field and its relative helicity  are given in Table~\ref{table1}.  To illustrate the computation for a given bounding radius $r_{\rm L}$, we assume that the transition to the linear field occurs with  $r_{\rm L} = 7.0\,R$. 
The linear field is given by the model of \citet{ChandrasekharKendall57} as introduced in Sect.~\ref{energy_star_field}. The value of the force-free parameter $\alpha$ is not known a priori and will be determined by the condition that the relative helicity of the linear field $H_{\rm R(LFF)}$ be equal to the helicity of the non-linear Wolfson field $H_{\rm R (NLLF)}$. 

To implement our method, we compute $H_{\rm R(LFF)}$ and the energy $E_{\rm LFF}$ of the linear field for 200 values of $\alpha$ in the range from $0.1$ to $1.1\, R^{-1}$, chosen to broadly  encompass the true unknown value. We truncate the series  of the  poloidal and toroidal components at the order $l = 50$ and consider only the odd values of $l$ given that the non-linear field is antisymmetric with respect to the equatorial plane of the star. For each order $ 1 \leq l \leq 50$ and a given $\alpha$, we solve Eqs.~(\ref{linear_field_coeffs}) with $p_{l}$ given by Eq.~(\ref{pl_coeff}), where $B_{r}(R,x) = -B_{0}f^{\prime}(x)$, to find $c_{l}$ and $d_{l}$. Then we use Eqs.~(\ref{poloidal_toroidal_comps}) and~(\ref{z_function}) to find the field components. 

From the toroidal component $B_{\phi}$, we evaluate the relative helicity by Eqs.~(\ref{helicity_linear}) and~(\ref{toroidal_energy}) numerically integrating the functions $r^{2} [Z_{l}(\alpha r)]^{2}$ in the interval $[R, r_{\rm L}]$. The energy of the linear field $E_{\rm LFF}$ is computed from Eq.~(\ref{linear_energy_eps}) where the surface terms $\epsilon_{l}$ come from Eq.~(\ref{epsilon_coeff}) after the surface field components have been computed from Eqs.~(\ref{poloidal_toroidal_comps}). 

\begin{figure}
\flushleft{ 
 \includegraphics[width=9cm,height=9cm]{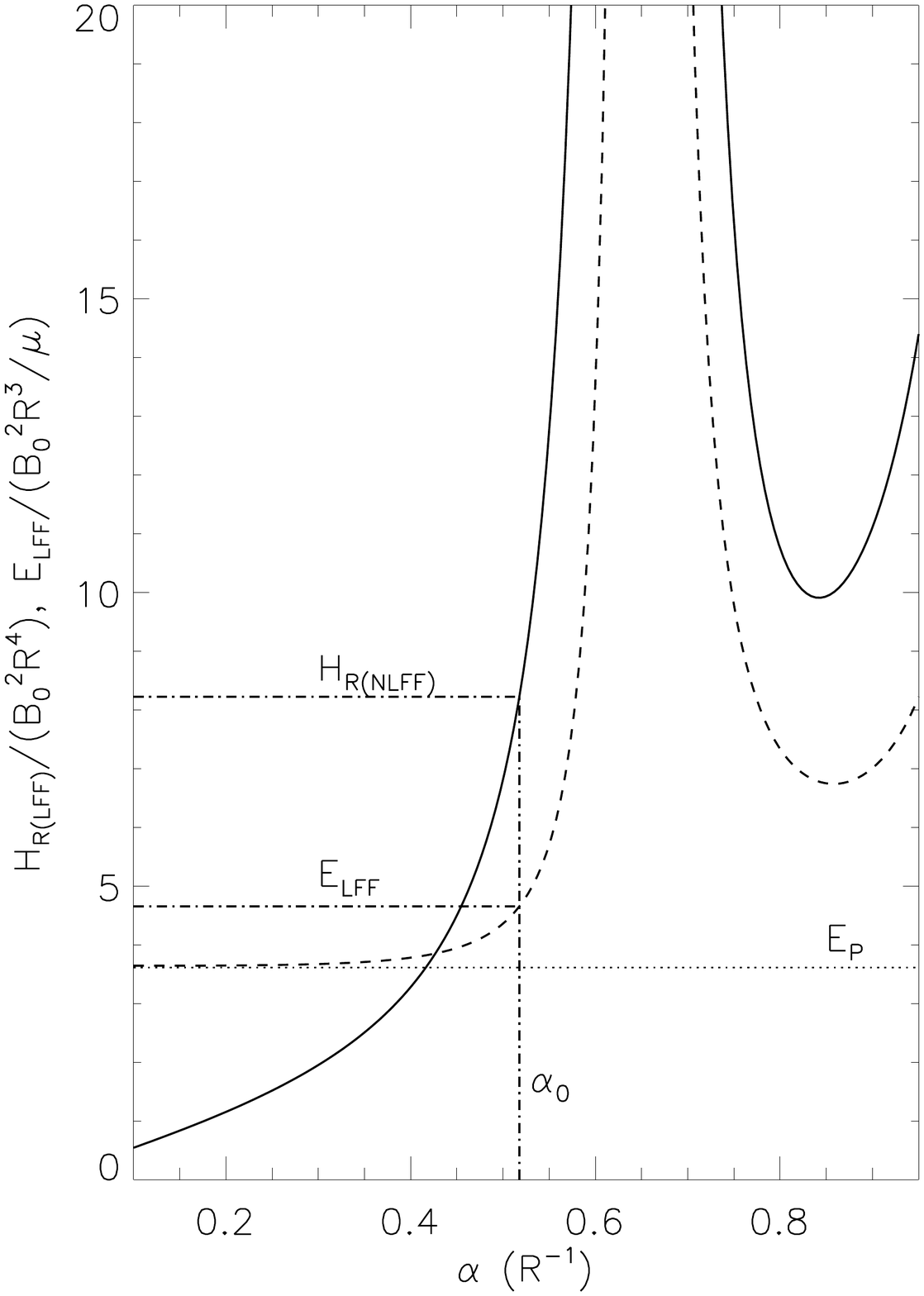}} 
   \caption{Relative helicity $H_{\rm R (LFF)}$ (solid line) and energy $E_{\rm LFF}$ (dashed line) of the linear force-free field within the magnetic boundary surface $r_{\rm L} = 7.0\,R$ versus the force-free parameter $\alpha$. The radial field at the photosphere $r=R$ is the same as for the non-linear field of \citet{Wolfson95} with $n=0.5$ discussed in Sect.~\ref{star_field}. The absolute minimum energy of the potential field with the same boundary conditions at the photosphere $E_{\rm p}$ is indicated by the horizontal dotted line. The  relative helicity $H_{\rm R(NLFF)}$ of the non-linear field is indicated by the horizontal dash-dotted line, while the vertical line with the same linestyle gives the corresponding value $\alpha_{0}$ of the parameter $\alpha$ and the energy $E_{\rm LFF}$ of the linear field with the same relative helicity as the non-linear field (see the text). }
              \label{Fig3}%
\end{figure}

The relative helicity and the magnetic energy are plotted vs. $\alpha$ in Fig.~\ref{Fig3} for $\alpha$ in the chosen range. It is so extended that the first value of $\alpha$ corresponding to the solution of the homogeneous boundary value problem, i.e., $B_{r}(R,x)=B_{r}(r_{\rm L}, x) = 0$, falls inside the interval as shown by the divergence of $H_{\rm R(LFF)}$ and $E_{\rm LFF}$ for $\alpha = \alpha_{\rm e} \sim 0.65\, R^{-1}$ \citep[cf. Fig.~1 in][]{Berger85}. Therefore, we consider only the first branch of the helicity plot (solid line) to find the value of $\alpha = \alpha_{\rm 0}$ that makes $H_{\rm R (LFF)} (\alpha_{0}) = H_{\rm R (NLFF)}$, thus obtaining $\alpha_{0} = 0.51769\, R^{-1}$. 

Having found $\alpha_{0}$, from the plot of $E_{\rm LFF}$ vs. $\alpha$ (dashed line), we determine the energy of the sought linear force-free field with the same photospheric boundary conditions, the outer magnetic surface at $r = r_{\rm L}$, and the same relative helicity of the considered Wolfson field. The energy of the linear field is $E_{\rm LFF} (\alpha_{0}) = 4.6534\, B_{0}^{2} R^{3}/{\mu}$. Therefore, the magnetic energy released in the transition that can power a stellar flare is: $E_{\rm NLFF} - E_{\rm LFF} (\alpha_{0}) = 0.2497\, B_{0}^{2} R^{3}/{\mu}$. 

\begin{figure*}
\flushleft{ 
 \includegraphics[width=17cm,height=11cm]{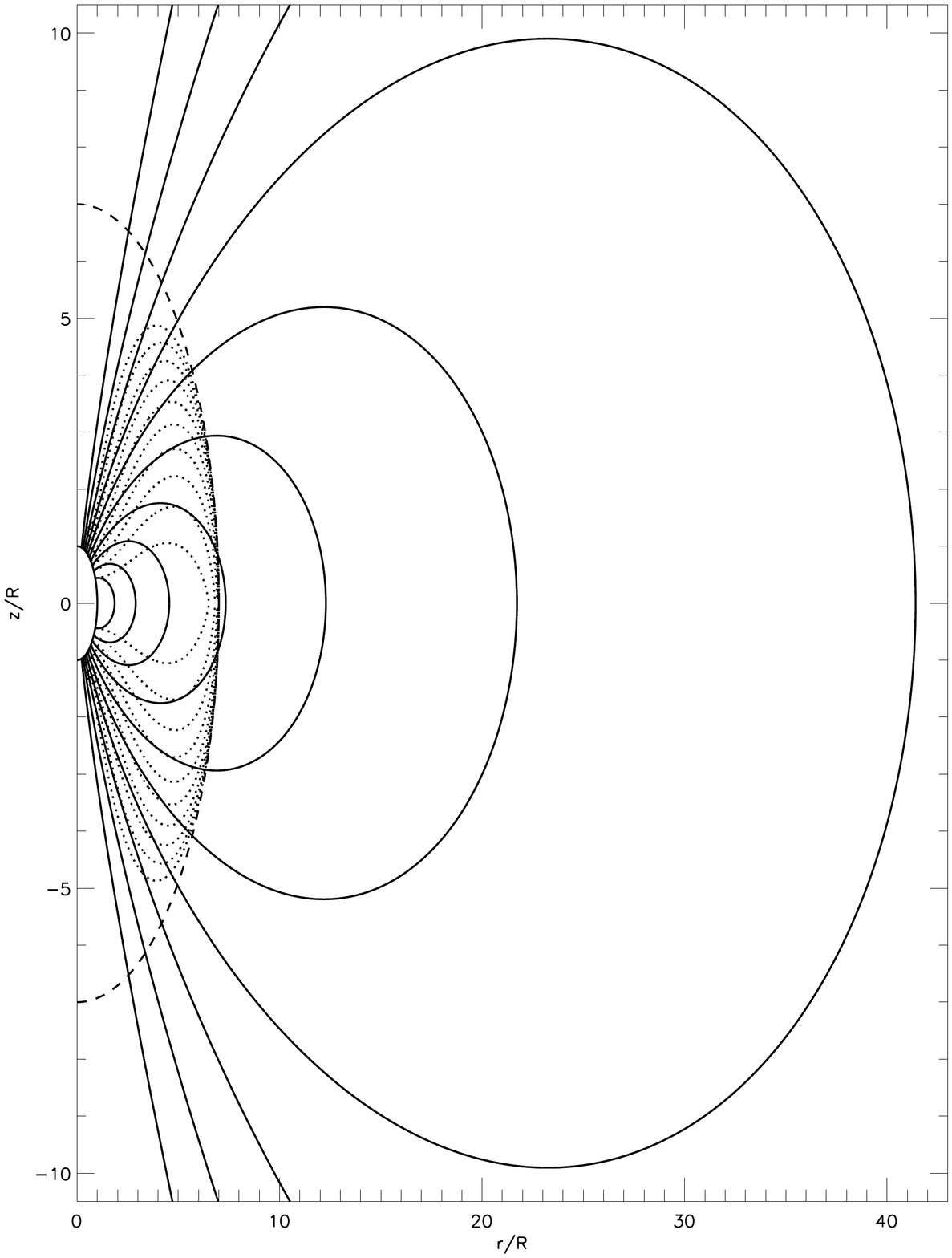}} 
   \caption{Projection of the magnetic field lines onto the meridional plane for the non-linear force-free model of \citet{Wolfson95} with $n=0.5$ as described in Sect.~\ref{star_field} (solid lines). The linear model with the same photospheric boundary conditions and magnetic helicity, confined within the radius $r_{\rm L}=7.0\,R$, is shown for comparison (dotted lines). The magnetic surface at $r = r_{\rm L}$ is also shown (dashed line). }
              \label{Fig4}%
\end{figure*}

In Fig.~\ref{Fig4}, we plot a meridional section of the field lines of the non-linear Wolfson field with $n=0.5$ and of the linear field with the same photospheric boundary conditions, magnetic bounding surface at $r=r_{\rm L} = 7.0\, R$, and $H_{\rm R(LFF)} = H_{\rm R(NLFF)}$. 

\begin{figure}
\flushleft{ 
 \includegraphics[width=9cm,height=8cm]{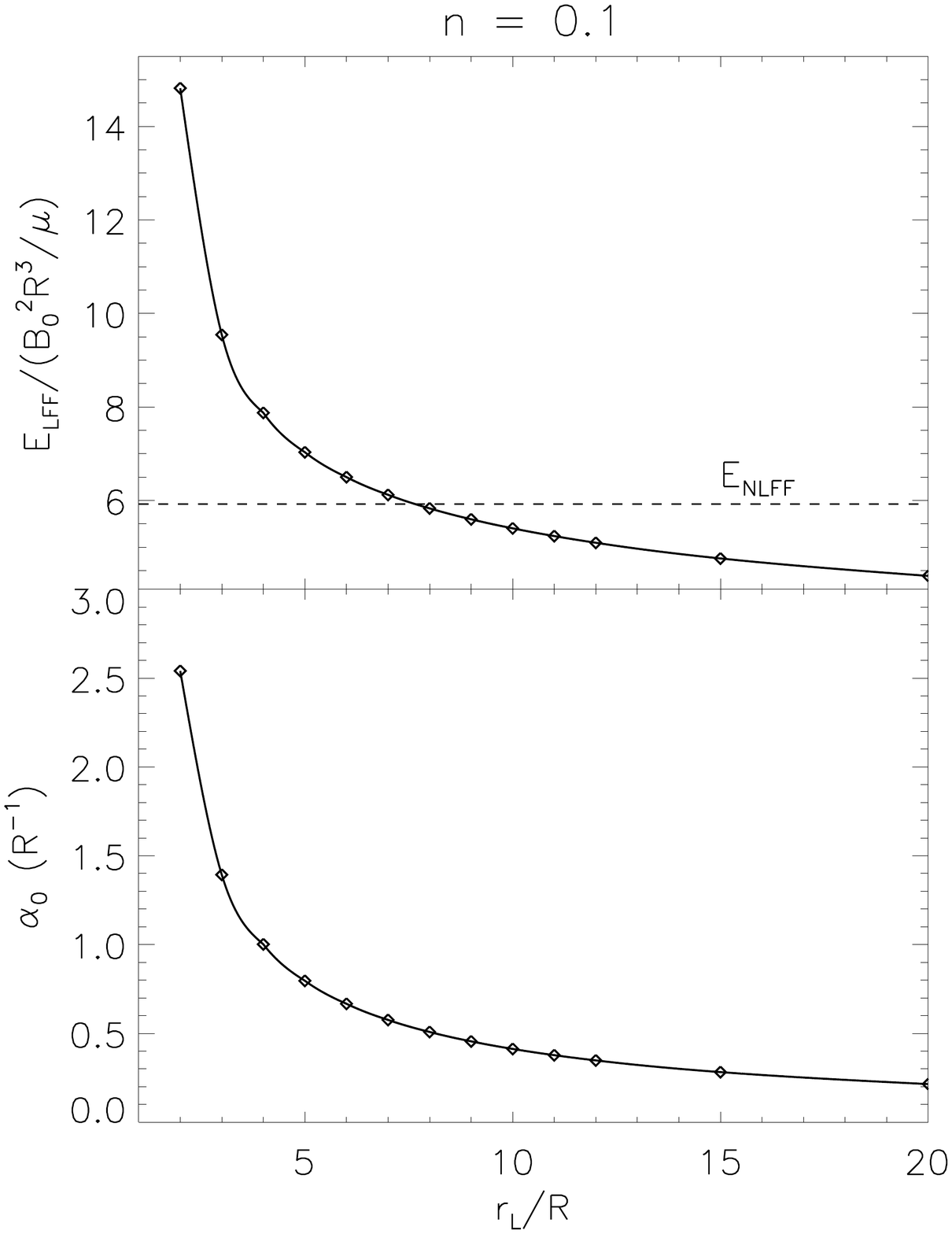}} 
   \caption{Top panel: Energy of the linear force-free field with outer boundary surface at $r_{\rm L}$ and the same relative helicity and boundary conditions on the stellar surface as the non-linear Wolfson field with $n=0.1$. The solid line is a spline interpolation through the computed values as indicated by the open diamonds. The energy of the non-linear field is indicated by the horizontal dashed line. Botton panel: the value of the force-free parameter $\alpha_{0}$ of the linear field satisfying the above conditions vs. the radius of the outer boundary surface.  The solid line is a spline interpolation through the computed values (open diamonds). }
              \label{Fig5}%
\end{figure}

\begin{figure}
\flushleft{ 
 \includegraphics[width=9cm,height=8cm]{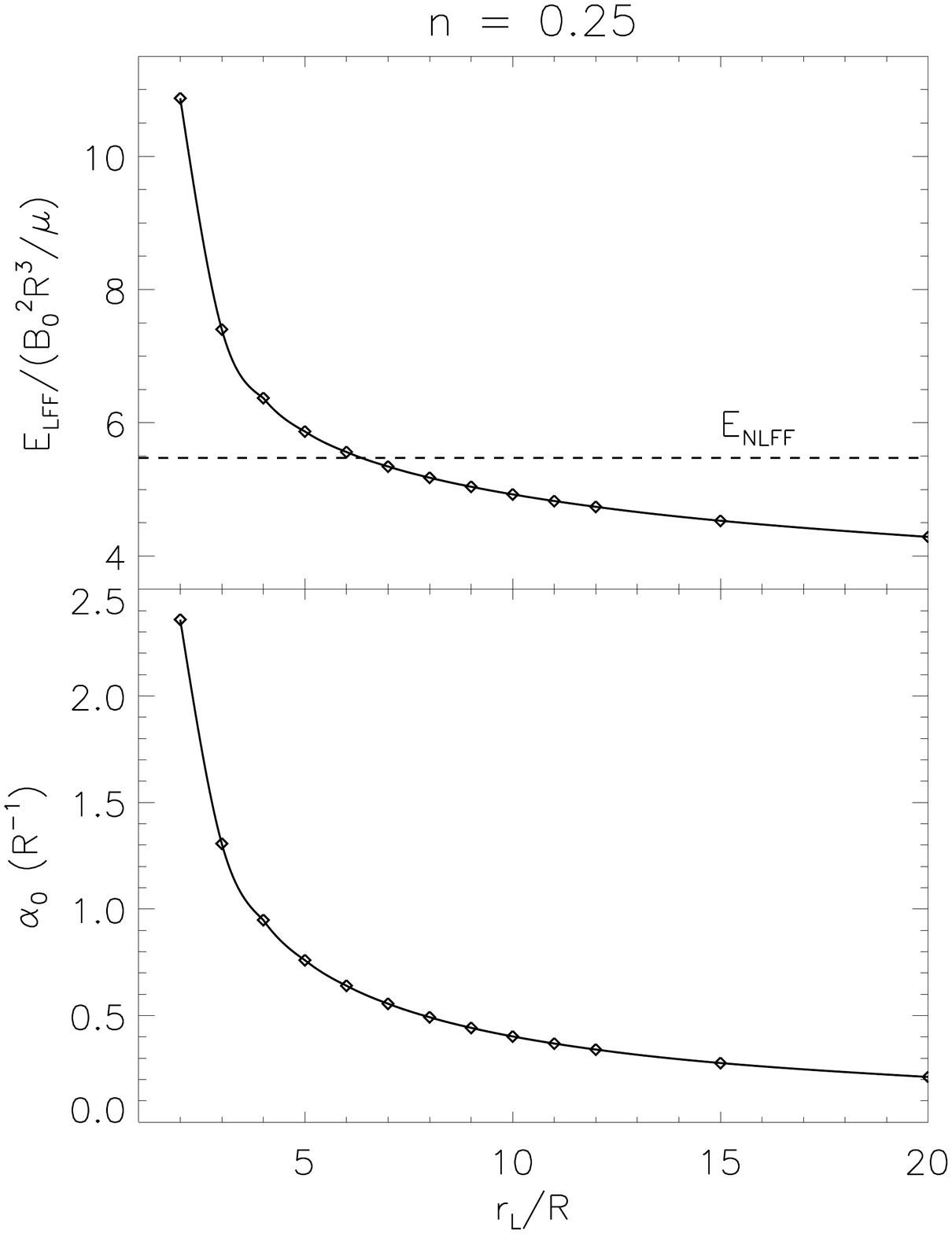}} 
   \caption{As of Fig.~\ref{Fig5}, for the Wolfson field with $n=0.25$.  }
              \label{Fig6}%
\end{figure}

\begin{figure}
\flushleft{ 
 \includegraphics[width=9cm,height=8cm]{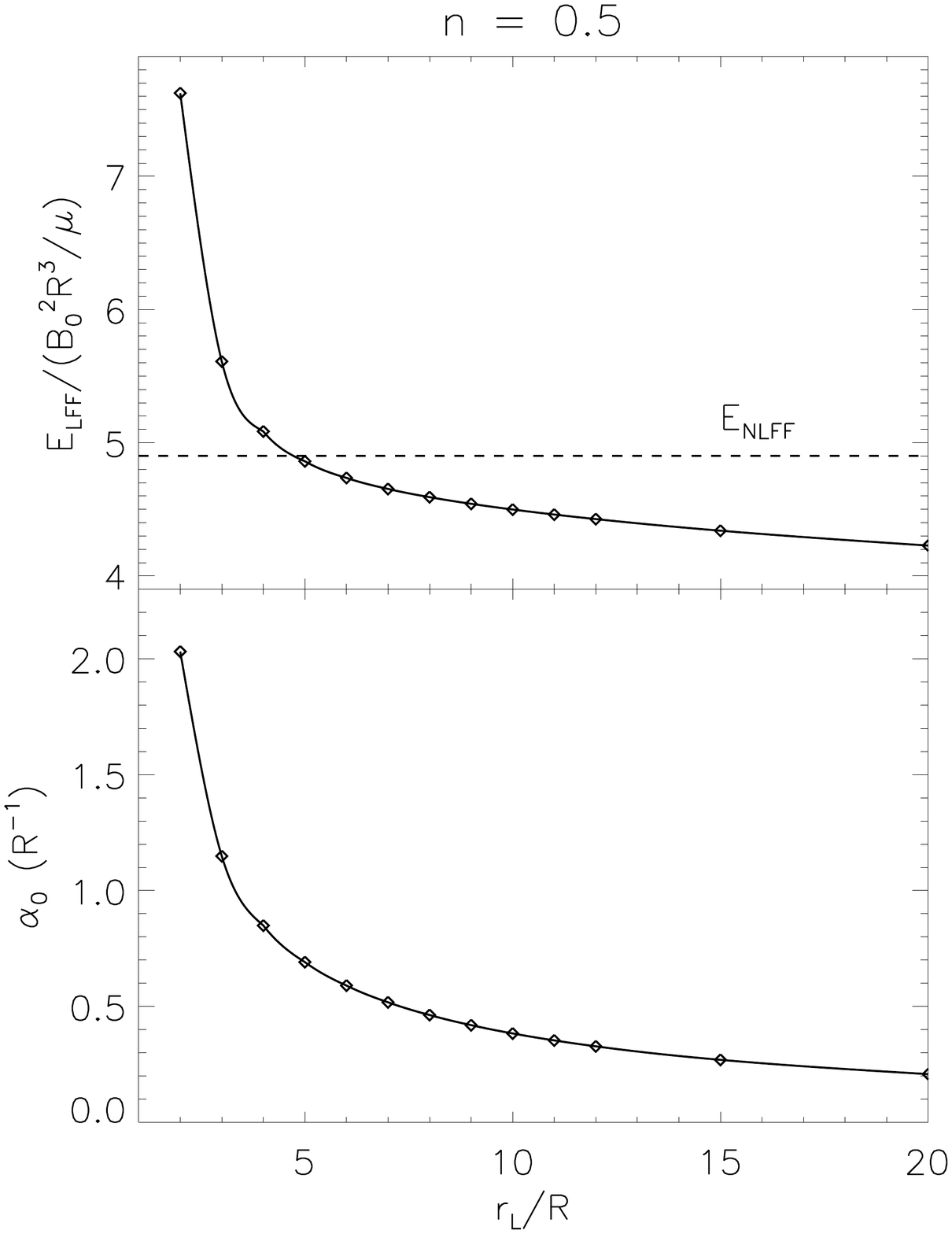}} 
   \caption{As of Fig.~\ref{Fig5}, for the Wolfson field with $n=0.5$.  }
              \label{Fig7}%
\end{figure}

The procedure outlined above for a specific value of $r_{\rm L}$ can be repeated for different values of the radius of the outer boundary surface, thus obtaining the corresponding value of the energy of the linear field $E_{\rm LFF}$ vs. $r_{\rm L}$. The results of these calculations are shown in Figs.~\ref{Fig5}, \ref{Fig6}, and~\ref{Fig7} for $n=0.1, 0.25$, and $0.5$, respectively.  We see that the energy of the linear field is greater than the energy of the non-linear field when the boundary radius $r_{\rm L}$ is smaller than a certain radius $r_{\rm E}$ that depends on the value of $n$ of the non-linear field.  This implies that the transition from the non-linear to the linear field can occur spontaneously only if $r_{\rm L} \geq r_{\rm E}$ as anticipated in Sect.~\ref{energy_star_field}. The transition  leads to a state of minimum energy dissipation rate only if the periastron distance of the planet is greater than $r_{\rm L}$. Therefore, we find that this state can be reached only if $a(1-e) \geq r_{\rm E}$. For the specific non-linear fields we considered, $r_{\rm E}$ ranges between $4.5\, R$ and $8.0\, R$, implying that very close-in planets produce a continuous energy dissipation through reconnection between their own magnetic fields and the stellar coronal field \citep[cf.][]{Lanza09,Lanza12}. Note that the value of $r_{\rm E}$ can be remarkably different for other non-linear field configurations. 

\begin{figure}
\centering{
 \includegraphics[width=9cm,height=10cm,angle=90]{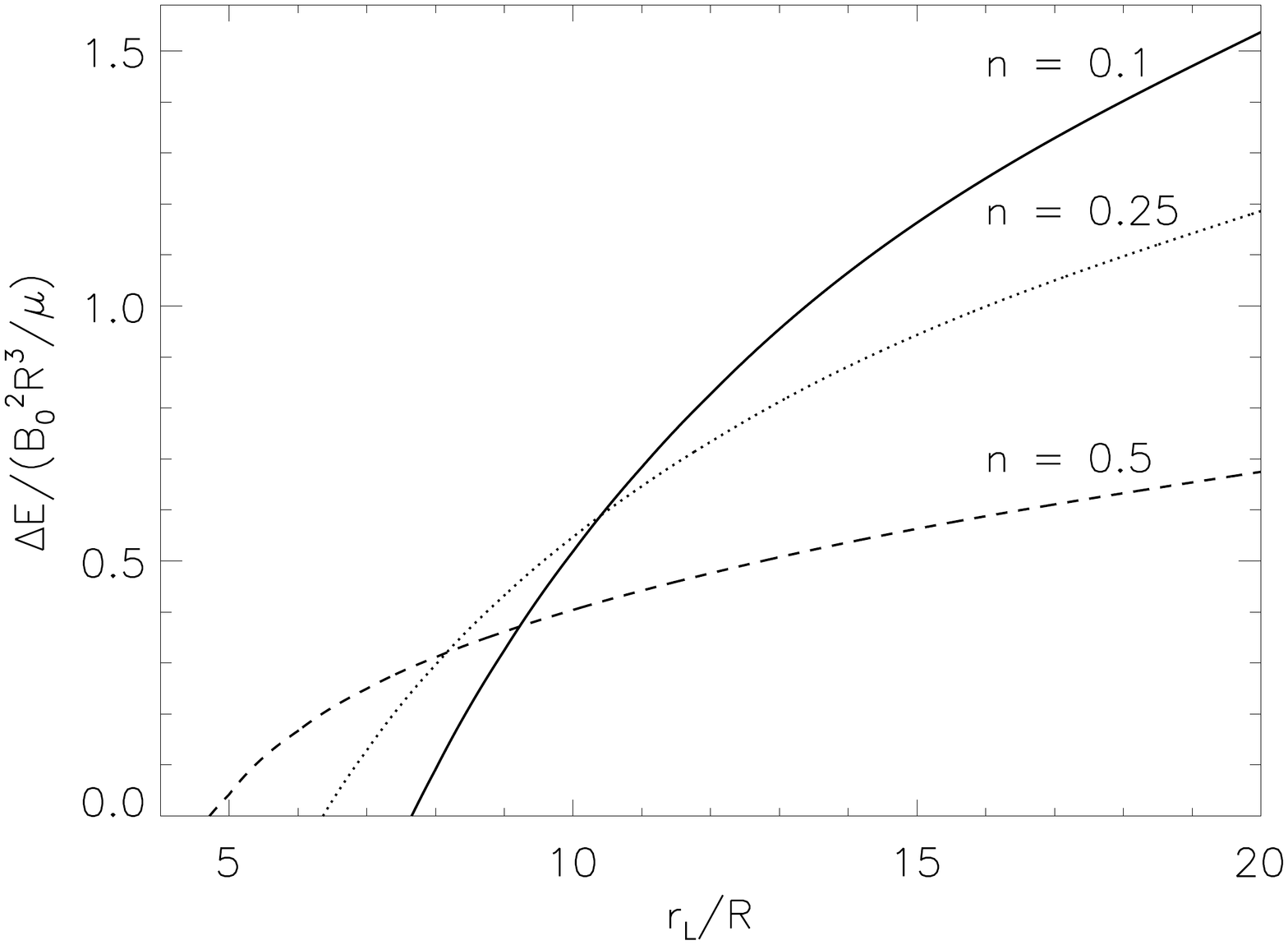}} 
   \caption{The magnetic energy $\Delta E$ released in the transition from the non-linear Wolfson field to the corresponding linear field vs. the boundary radius $r_{\rm L}$ of the latter. Different linestyles indicate the different values of the index $n$ of the Wolfson field as labelled.   }
              \label{free_energy}%
\end{figure}

When $r_{\rm L} \geq r_{\rm E}$, the amount of energy that is released in the transition from the Wolfson field to the linear field with the same relative helicity and photospheric boundary conditions is plotted in Fig.~\ref{free_energy} vs. the radius $r_{\rm L}$ of the outer boundary surface and for different values of the parameter $n$. For a young star with a strong magnetic field and a corona with an extension of $8-12\,R$, the amount of energy that can be released to power a large flare is between $\sim 0.3$ and $\sim 0.8\,B_{0}^{2} R^{3}/\mu$. This is comparable to the energy released in the transition from the open Aly field to the minimum energy potential field in the case of the non-linear fields considered by \citet{Flyeretal04}. 

\begin{figure}
\centering{
 \includegraphics[width=9cm,height=10cm,angle=90]{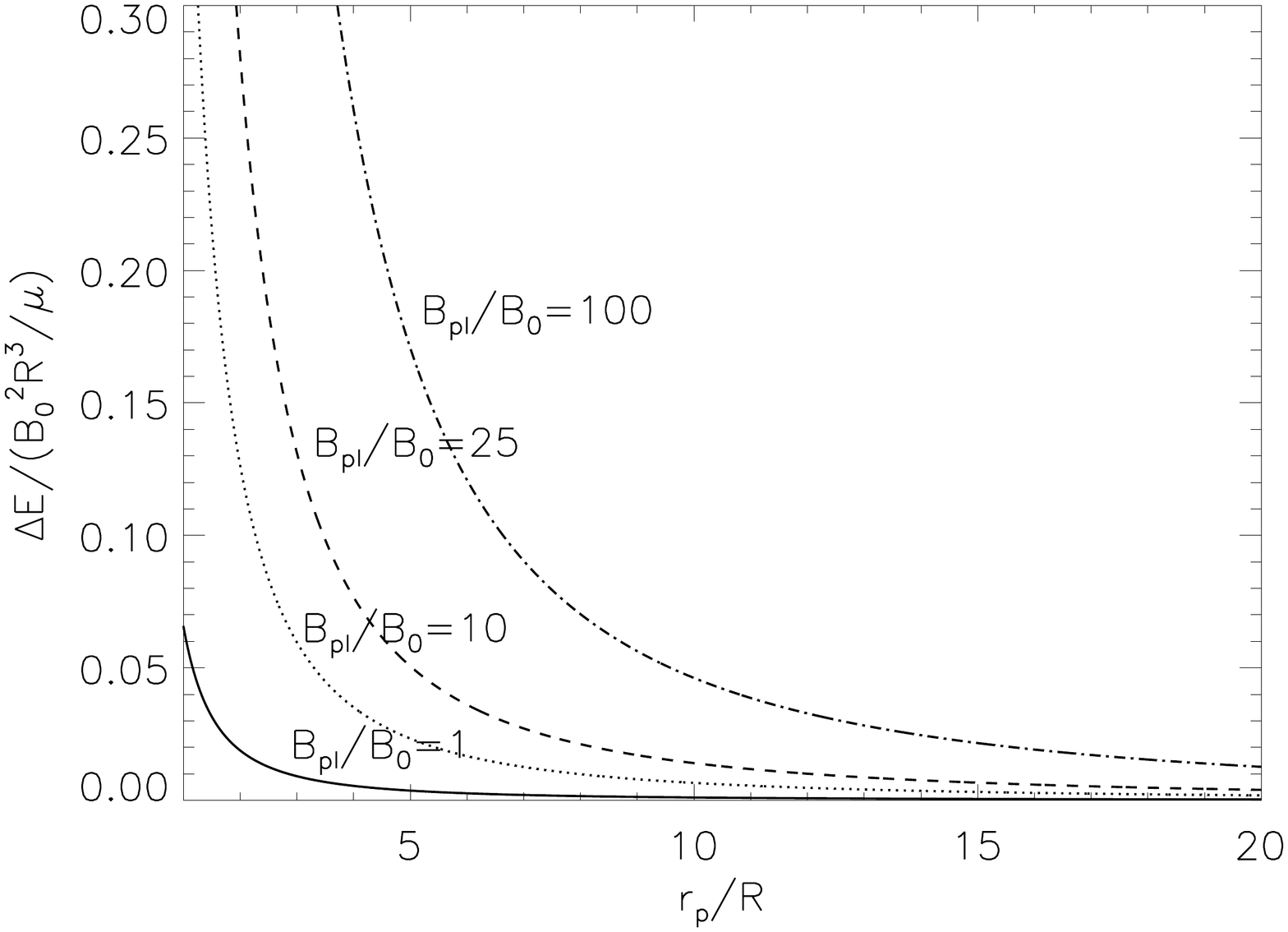}} 
   \caption{Magnetic energy difference $\Delta E$ between the stellar coronal field without and with a planetary magnetosphere inside itself vs. the distance $r_{\rm p}$ of the planet from the centre of its host star. The value of $\Delta E$ is evaluated in the case of a closed magnetosphere ($C_{\rm imf} = 0$) and a potential stellar field  directed radially by applying Eq.~(\ref{energy_var_magsp}) with $n = 0$. Different linestyles refer to different value of the ratio of the planetary to the stellar magnetic field $B_{\rm pl}/B_{0}$ as labelled. The radius of the planet is assumed to be $0.1\, R$ in all the cases. } 
              \label{free_energy_radial_field}%
\end{figure}

Considering now the case of older, weakly or moderately active stars, we evaluate the energy $\Delta E$ made available when the planetary magnetosphere perturbs the stellar coronal field that we assume to be radially directed away from the star. We make use of Eqs.~(\ref{rm_eq}) and~(\ref{energy_var_magsp}) considering the case of a closed magnetosphere ($C_{\rm imf} = 0$) and a radial field ($\cos \xi = 0$) because these maximize the energy perturbation. In Fig.~\ref{free_energy_radial_field}, $\Delta E$ is plotted vs. the distance of the planet $r_{\rm p}$ for different values of the ratio between the planetary and the stellar field intensities $B_{\rm pl}/B_{0}$ and considering a typical hot Jupiter with a radius $R_{\rm pl} = 0.1\, R$. 

The magnetic fields of hot Jupiters are highly uncertain because only very indirect measures have been obtained so far ranging from $5-10$~G to $20-25$~G \citep[][]{Vidottoetal10,Cauleyetal15,Rogers17}. Fields up to $\sim 100$~G are predicted by theoretical dynamo models, in particular for massive planets ($M_{\rm pl}\geq 5\, M_{\rm Jup}$) with ages younger than $\approx 1-2$~Gyr \citep{ReinersChristensen10}. Given that the stellar magnetic field $B_{0}$ ranges between $\sim 1$ and $\sim 10$~G, we plot the energy differences for $B_{\rm pl}/B_{0}$ between 1 and 100.  

The relative energy variations induced by a planetary magnetosphere are typically $1-2$ orders of magnitude smaller than those produced by the other mechanisms discussed in the case of active stars. However, while those mechanisms involve a global transition of the stellar coronal field, the energy release induced by the magnetosphere can occur inside a radial field structure large enough to embed the entire magnetosphere at the distance of the planet. 
A large coronal streamer can be perturbed when the planet passes through it during its orbital motion triggering the release of the energy $\Delta E$ because the energy of the configuration with the embedded magnetosphere is lower than the energy of the streamer without it. 



\section{Applications}
\label{applications}

To illustrate the application of the results in Sect.~\ref{results} to specific planetary systems, eight representative cases have been selected. Their names and relevant parameters are given in Table~\ref{table_system_par} where we list, from the left to the right, the system name, the orbital period, the semimajor axis, the orbital eccentricity, the mass and the radius of the planet, the mass and the radius of the star, the surface magnetic field, and the reference for the field and  the rotation period. We selected HD~17156 because it has a remarkably eccentric orbit and showed flare activity at the periastron \citep{Maggioetal15}; HD~80606 was suggested as a promising target for the observation of enhanced activity, including flaring, at the periastron, but the observations have not been able to reveal them yet \citep{Figueiraetal16}; HD~189733 is a well-studied system that exhibited repeated flaring {  possibly at a preferential orbital phase} \citep{Pillitterietal14,Pillitterietal15}; HD~179949 and $\tau$~Bootis are F-type stars with hot Jupiters {  for which a star-planet interaction leading to chromospheric and photospheric hot spots has been claimed}, respectively \citep{Shkolniketal05,Walkeretal08}; V830~Tauri and TAP~26 are weak-line T~Tauri stars with very strong surface magnetic fields accompanied by hot Jupiters \citep{Donatietal17,Yuetal17} making them unique targets to study star-planet interactions in very young systems because their ages are $\sim 2$~and $\sim 17$~Myr, respectively; finally, Kepler-78 is an example of a very close-in planet around a late-type star whose magnetic field has been measured by \citet{Moutouetal16}. {  Note that conclusive evidence of magnetic star-planet interaction is not available yet for these as well as for other systems. A critical discussion of the existing observations is beyond the scope of the present work and can be found elsewhere \citep[e.g.][]{Shkolniketal08,Milleretal15}. We consider these systems because they can be useful testbeds for our models and the previous claims of star-planet interactions for some of them make them interesting for future observational campaigns. In particular, an estimate of the energy available to produce flares in their stars can provide guidance for those future observations. }

For HD~17156 and HD~80606, the system parameters were extracted from the exoplanets.org database~~\citep{Hanetal14}, while the maximum intensity of the stellar surface field was guessed considering  hosts with the same spectral type and rotation period \citep[cf.][]{Faresetal13}. The rotation period was estimated from the projected rotation velocity $v \sin i$ and stellar radius. Since these are transiting systems, their projected obliquity $\lambda$ was measured yielding values of $ 10^{\circ} \pm 5^{\circ}$ for HD~17156 and $42^{\circ} \pm 8^{\circ}$ for HD~80606. We approximated $i \approx 90^{\circ}- \lambda$ to estimate $P_{\rm rot}$ for these systems. 

For the other hosts, the parameters were extracted from exoplanets.org, while the maximum magnetic field was measured through spectropolarimetric techniques and we give  the corresponding reference in Table~\ref{table_system_par}.  The rotation period is given in the same reference, often together with information on surface differential rotation, because a detailed knowledge of stellar rotation is needed to reconstruct maps of the photospheric magnetic fields. The radius of the planets that do not transit their star (i.e., HD~179949, $\tau$~Boo, V830~Tau, and TAP~26) is assumed to be equal to the radius of Jupiter. 
\begin{table*}
\caption{Parameters of the planetary systems chosen to illustrate the application of the present theory.}
\begin{center}
\begin{tabular}{lrccrrccrrc}
\hline
 System name &   $P_{\rm orb}$ &  $a$  &  $e$  &   $M_{\rm pl}$  &    $R_{\rm pl}$ &  $M$ &  $R$ & $B_{0}$ & $P_{\rm rot}$ & Reference for $B_{0}$ and $P_{\rm rot}$ \\
   & (d)~ & (AU) & & ($M_{\rm J}$) & ($R_{\rm J}$) & ($M_{\odot}$) & ($R_{\odot}$) & (G) & (d) & \\
   \hline
        HD~17156 &    21.2166 &     0.1632 &      0.682 &   3.30 &   1.02 &   1.28 &   1.51 &    3.0 &     27.2  & estimated \\
        HD~80606 &   111.4370 &     0.4473 &      0.934 &   3.89 &   1.03 &   0.96 &   0.98 &   10.0 &     18.4  & estimated \\
       HD~189733 &     2.2186 &     0.0310 &      0.0 &   1.14 &   1.14 &   0.81 &   0.76 &   36.0 &     11.94  & \citet{Faresetal10} \\
       HD~179949 &     3.0925 &     0.0439 &      0.0 &   0.90 &   1.00 &   1.18 &   1.23 &    3.7 &      7.62  & \citet{Faresetal12} \\
         $\tau$ Bootis &     3.3124 &     0.0490 &      0.0 &   5.95 &   1.00 &   1.39 &   1.42 &    3.9 &      3.14  & \citet{Mengeletal16} \\
        V830~Tauri &     4.9270 &     0.0570 &      0.0 &   0.70 &   1.00 &   1.00 &   2.00 &  350.0 &      2.741  & \citet{Donatietal17}\\
          TAP~26 &    10.7900 &     0.0968 &      0.0 &   2.03 &   1.00 &   1.04 &   1.17 &  120.0 &      0.714  & \citet{Yuetal17}\\
       Kepler~78 &     0.3550 &     0.0092 &      0.0 &   0.01 &   0.10 &   0.81 &   0.74 &   16.0 &     12.59  & \citet{Moutouetal16}\\
       \hline
\end{tabular}
\end{center}
\label{table_system_par}
\end{table*}

In our model, the radius $r_{\rm L}$ of the closed corona is the minimum between the radius where the parameter $\beta = 1$ on the equatorial plane of the star and the periastron distance of the planet. To compute $\beta (r)$ and the Alfven velocity, we assume an electron density at the base of the corona $n_{\rm e} = 10^{14}$~m$^{-3}$ and a temperature $ T = 10^{6}$~K for all our stars and apply Eq.~(\ref{beta_vs_radius}) with the magnetic field strength as given by the Wolfson field for $B_{0}$ given in Table~\ref{table_system_par} and consider the three values of the parameter $n = 0.1$, $0.25$, and $0.5$. We choose these relatively low temperature and base density to have a closed corona ($\beta < 1$) extending up to the periastron distance of our planets. On the other hand, if we assume $T = 10^{7}$~K with the same base density, only the very active stars V830~Tau and TAP~26 will have a closed corona extending up to the distance of their planets \citep[cf. Sect.~3.1 of][]{Lanza12}. 

In Table~\ref{close_corona_limit}, we list, from the left to the right, the name of the system, the parameter $n$, the maximum radial extension of the closed corona $r_{\rm L}$, the Alfven transit time from the surface of the star to the limit of the closed corona $\tau_{\rm A} = \int_{R}^{r_{\rm L}} [v_{\rm A}(r)]^{-1}\, dr$, the maximum energy available in the case of the Wolfson field $\Delta E_{\rm max} ({\rm W})$ as well as in the case of the fields of \citet{Flyeretal04} $\Delta E_{\rm max} ({\rm F})$. 

The maximum extension of the closed corona $r_{\rm L}$ is given by the periastron distance of the planets in all the considered systems.  
The radial Alfven transit time across the closed corona $\tau_{\rm A}$ is of the order of $10^{2}-10^{3}$~s in the case of weakly or moderately  active stars and of the order of $10-10^{2}$~s in the case of the most active stars. Note that  $\tau_{\rm A}$ would increase for  coronae with a higher base density or temperature. 

The maximum available magnetic energy $\Delta E_{\rm max}$ is computed as the difference between the Aly energy and the energy of the potential magnetic field with the same photospheric boundary conditions. In the case of the fields of Flyer et al., these are independent of $n$ because their photospheric boundary conditions are those of a potential dipole in all the cases. We recall that $\Delta E_{\rm max}$ is available if the field has enough energy to open up all its field lines and get rid of all of its helicity. This is possible in the case of the fields by Flyer et al. by a continuous accumulation of helicity, but requires some additional source of energy in the case of the Wolfson field. In the case of HD~17156, $\Delta E_{\rm max}$  is comparable with the energy released in the largest solar flares ever observed as expected given that this star has values of $B_{0}$ and $R$ similar to those of the Sun. For the other stars, $\Delta E_{\rm max}$ is larger by 1 to 4.5 orders of magnitude, mainly because of the stronger surface field $B_{0}$ (cf. Table~\ref{table_system_par}). 

In the case of the selected systems, $r_{\rm L}  > r_{\rm E}$ for the Wolfson fields considered in Sect.~\ref{results}, with a few exceptions. {  Therefore, we can consider the transition from the non-linear Wolfson field to a confined linear field and apply the theory in Sect.~\ref{energy_star_field} and the results in Fig.~\ref{free_energy}.  In this way, we obtain the energy values listed in Table~\ref{non-linear-linear}. }There we report, from the left to the right, the name of the system, the parameter $n$ of the non-linear Wolfson field, the energy $\Delta E$ released in the transition from the non-linear to the linear field with the same photospheric boundary conditions and relative helicity (if positive), and the mean available power computed as $\Delta E/\tau_{\rm A}$, where $\tau_{\rm A}$ is taken from Table~\ref{close_corona_limit}. The free magnetic energy $\Delta E$ is significantly lower than the maximum values $\Delta E_{\rm max}$ in Table~\ref{close_corona_limit}, but it is still more than enough to account for the energy of the flares observed in HD~17156  or HD~189733. Specifically, in the former case, \citet{Maggioetal15} estimate an emitted power of $5 \times 10^{19}$~W in the X-rays, while in the latter case \citet{Pillitterietal14} estimate a total flare energy of the order of $10^{25}$~J.  In the case of HD~179949, the mean power is comparable or larger than that emitted by the chromospheric hot spots {  possibly} associated with the planet, estimated to be of $\approx 10^{20}$~W according to \citet{Shkolniketal05}. 
An individual flare was observed in the X-rays by \citet{Scandariatoetal13} with an estimated energy of $\approx 5 \times 10^{24}$~J that is of the same order of magnitude of the magnetic energy available in HD~179949. 
In the case of the young stars V830~Tauri and TAP~26, the available energy is $2-3$ orders of magnitude larger than in the most powerful solar flares, that is $\sim 10^{28}-10^{29}$~J. For  Kepler-78, we cannot invoke our model to compute the available energy because its planet is so close that $r_{\rm L} < r_{\rm E}$, thus the considered field transition cannot release energy.

\begin{table*}
\begin{center}
\caption{Radial extension of the closed isothermal corona, Alfven transit time, and maximum available magnetic energy for our systems  with $n_{\rm e}= 10^{14}$~m$^{-3}$ and $T = 10^{6}$~K.}
\begin{tabular}{lcrccc}
\hline
  System name  &  $n$ &        $r_{\rm L}$~ &            $\tau_{\rm A}$ &  $\Delta E_{\rm max} ({\rm W}) $  & $\Delta E_{\rm max} ({\rm F})$
             \\
            & & (R) & (s) & (J) & (J) \\
            \hline
        HD 17156 & 0.10 &        7.400 &    3.097e+03 &    2.808e+26 &    5.728e+25  \\
                & 0.25 &        7.400 &    3.778e+03 &    2.629e+26 &    5.728e+25  \\
                & 0.50 &        7.400 &    5.352e+03 &    2.464e+26 &    5.728e+25  \\
        HD 80606 & 0.10 &        6.471 &    2.957e+02 &    8.580e+26 &    1.750e+26  \\
                & 0.25 &        6.471 &    3.459e+02 &    8.034e+26 &    1.750e+26  \\
                & 0.50 &        6.471 &    4.574e+02 &    7.528e+26 &    1.750e+26  \\
       HD 189733 & 0.10 &        8.808 &    6.569e+01 &    5.105e+27 &    1.041e+27  \\
                & 0.25 &        8.808 &    7.975e+01 &    4.780e+27 &    1.041e+27  \\
                & 0.50 &        8.808 &    1.131e+02 &    4.479e+27 &    1.041e+27  \\
       HD 179949 & 0.10 &        7.695 &    1.304e+03 &    2.300e+26 &    4.692e+25  \\
                & 0.25 &        7.695 &    1.572e+03 &    2.153e+26 &    4.692e+25  \\
                & 0.50 &        7.695 &    2.191e+03 &    2.018e+26 &    4.692e+25  \\
         $\tau$ Bootis & 0.10 &        7.412 &    1.110e+03 &    3.970e+26 &    8.099e+25  \\
                & 0.25 &        7.412 &    1.309e+03 &    3.717e+26 &    8.099e+25  \\
                & 0.50 &        7.412 &    1.760e+03 &    3.484e+26 &    8.099e+25  \\
        V830 Tauri & 0.10 &        6.122 &    8.549e+01 &    8.934e+30 &    1.822e+30  \\
                & 0.25 &        6.122 &    1.030e+02 &    8.365e+30 &    1.822e+30  \\
                & 0.50 &        6.122 &    1.420e+02 &    7.839e+30 &    1.822e+30  \\
          TAP 26 & 0.10 &       17.771 &    1.613e+01 &    2.102e+29 &    4.289e+28  \\
                & 0.25 &       17.771 &    1.755e+01 &    1.969e+29 &    4.289e+28  \\
                & 0.50 &       17.771 &    2.043e+01 &    1.845e+29 &    4.289e+28  \\
       Kepler 78 & 0.10 &        2.656 &    4.982e+01 &    9.457e+26 &    1.929e+26  \\
                & 0.25 &        2.656 &    5.254e+01 &    8.855e+26 &    1.929e+26  \\
                & 0.50 &        2.656 &    5.764e+01 &    8.298e+26 &    1.929e+26  \\
       \hline
\end{tabular}
\label{close_corona_limit}
\end{center}
\end{table*}

\begin{table}
\begin{center}
\caption{Energy released in the transition from the non-linear Wolfson field to the linear confined field with $n_{\rm e}= 10^{14}$~m$^{-3}$ and $T = 10^{6}$~K. }
\begin{tabular}{lccc}
\hline
   System name  &  $n$  &   $\Delta E$  &  $\Delta E/\tau_{\rm A}$        \\
                    &  & (J) & (W)  \\
           \hline
        HD 17156 & 0.10 &          < 0 &            -   \\
                & 0.25 &   1.6643e+25 &   4.4053e+21   \\
                & 0.50 &   2.2872e+25 &   4.2734e+21   \\
        HD 80606 & 0.10 &          < 0 &            -   \\
                & 0.25 &   5.9791e+24 &   1.7288e+22   \\
                & 0.50 &   5.3181e+25 &   1.1626e+23   \\
       HD 189733 & 0.10 &   4.2689e+26 &   6.4985e+24   \\
                & 0.25 &   6.1350e+26 &   7.6930e+24   \\
                & 0.50 &   5.2962e+26 &   4.6846e+24   \\
       HD 179949 & 0.10 &   7.7896e+23 &   5.9732e+20   \\
                & 0.25 &   1.6910e+25 &   1.0759e+22   \\
                & 0.50 &   1.9935e+25 &   9.1002e+21   \\
         $\tau$ Bootis & 0.10 &          < 0 &            -   \\
                & 0.25 &   2.3783e+25 &   1.8170e+22   \\
                & 0.50 &   3.2431e+25 &   1.8427e+22   \\
        V830 Tauri & 0.10 &          < 0 &            -   \\
                & 0.25 &          < 0 &            -   \\
                & 0.50 &   4.7140e+29 &   3.3200e+27   \\
          TAP 26 & 0.10 &   8.5732e+28 &   5.3151e+27   \\
                & 0.25 &   6.7236e+28 &   3.8307e+27   \\
                & 0.50 &   3.8864e+28 &   1.9028e+27   \\
       Kepler 78 & 0.10 &          < 0 &            -   \\
                & 0.25 &          < 0 &            -   \\
                & 0.50 &          < 0 &            -   \\  
     \hline 
\end{tabular}
\end{center}
\label{non-linear-linear}
\end{table}

When the temperature of the corona is greater than $(2-3) \times 10^{6}$~K or its base density is significantly higher than the value assumed above, $\, \beta > 1$ for our moderately active stars at the distance of their planets, so they orbit in the coronal region with open and radial field lines.  
In this case, we apply the theory in Sect.~\ref{magnetospheric_effect} (cf. also Fig.~\ref{free_energy_radial_field}) to compute the energy made available by the perturbation produced by the planetary magnetosphere. The results for our systems are listed in Table~\ref{magnetosphere} where we report, from the left to the right, the name of the system, the energy released in the interaction $\Delta E$, the magnetospheric crossing time $\tau_{\rm r}$ that is a measure of the timescale for energy release, and  the mean power available. The energy $\Delta E$  is computed by means of Eq.~(\ref{energy_var_magsp}) with $C_{\rm imf} = 0$ and $\xi = \pm\, \pi/2$ corresponding to a closed magnetosphere and a radially directed stellar field to maximize  the variation. The magnetic field of the planet is assumed to be $B_{\rm pl} = 100$~G also to maximize the effect. The crossing timescale comes from Eq.~(\ref{release_time}). {  The assumption that $P_{\rm rot} \gg P_{\rm orb}$ is not always verified for our systems (cf. Table~\ref{table_system_par}), but this changes the value of $\tau_{\rm r}$ only by a factor $\leq 2$, except in the case of $\tau$~Boo and TAP~26. Therefore, it  does not significantly affect the estimated powers given the uncertainties in the values of the stellar and planetary fields. }

In general,  the energy decrease due to the perturbation by the planet magnetosphere is between one and three orders of magnitude lower than the energy released by the previous mechanisms (cf. Tables~\ref{non-linear-linear} and~\ref{magnetosphere}). Therefore, the latter mechanisms dominate over the former  when they can operate. 

The mean power produced by the magnetospheric perturbation in the case of HD~17156 with the parameters in Table~\ref{table_system_par} can account for the flare observed by \citet{Maggioetal15}  in the X-rays. If we assume a surface field of $B_{0} = 1$~G, as those authors, the energy released, the crossing timescale, and the available power become $\Delta E = 2.841 \times 10^{23}$~J, $\tau_{\rm r}=3.515 \times 10^{4}$~s, and $8.084 \times 10^{18}$~W, that are insufficient to account for the observations.  Therefore, the models assuming a global transition of the coronal field are preferred in this case. In the case of HD~189733, we still find a total energy of the order of $10^{25}$~J with the present model, similar to the energy of the flares observed by \citet{Pillitterietal14}. 
For the chromospheric hot spots of HD~179949, \citet{Shkolniketal05} estimated an emitted power of the order of $10^{20}$~W that is a factor of three greater than our estimate in Table~\ref{magnetosphere}. Considering a stellar field $B_{0} = 8$~G, the energy increases by a factor of $\sim 2.5$ and gives a power comparable with the observations.  A field $B_{0} \sim 15$~G is required to account for the energy of the flare observed by \citet{Scandariatoetal13}. 

Finally, we note that in the case of the very close-by telluric planet Kepler-78, the released energy  is considerably lower because the volume of its magnetosphere is significantly smaller than in the case of the hot Jupiters owing to its smaller radius, closer distance, and relatively strong stellar field. However, the available power is comparable with that of the hot spots of HD~179949 making this system an interesting target to look for a similar phenomenon.

\begin{table}
\caption{Energy released, crossing timescale, and mean available power in the interaction between  the stellar coronal fields and the magnetospheres of our planets. }
\begin{center}
\begin{tabular}{lccc}
\hline
  System name &      $\Delta E$ &   $\tau_{\rm r}$ &     $\Delta E/ \tau_{\rm r}$  \\
          & (J) & (s) & (W) \\
\hline
       HD 17156 &      1.0159e+24 &       2.437e+04 &       4.168e+19  \\
        HD 80606 &      4.3575e+24 &       1.225e+04 &       3.557e+20  \\
       HD 189733 &      1.2519e+25 &       1.689e+04 &       7.414e+20  \\
       HD 179949 &      1.0353e+24 &       2.848e+04 &       3.635e+19  \\
        $\tau$ Bootis &      1.2725e+24 &       2.621e+04 &       4.856e+19  \\
        V830 Tauri &      6.0920e+26 &       6.588e+03 &       9.247e+22  \\
          TAP 26 &      1.5789e+25 &       2.470e+04 &       6.391e+20  \\
       Kepler 78 &      1.5777e+23 &       4.882e+02 &       3.232e+20  \\
\hline 
\end{tabular}
\end{center}
\label{magnetosphere}
\end{table}

\section{Discussion and conclusions}
{  Star-planet magnetic interactions are expected to produce stellar flares. Although the observational evidence is not conclusive yet, it is interesting to theoretically investigate possible mechanisms that can provide energy for such flares and estimate the maximum amount they can deliver.
In the present study, we investigated three different mechanisms that can operate in stars with different levels of magnetic activity.}

The first mechanism assumes that the energy of the large-scale stellar field is steadily increased together with its magnetic helicity by the emergence of new magnetic flux from the interior of the star, while the planet acts simply as a trigger to produce the flare when the accumulated helicity gets close to a threshold value.  This mechanism can operate independently of the presence of any planet, the difference in the case of star-planet interaction being the triggering action of the planet.  The mechanism considers a transition between an open field configuration having the so-called Aly energy and the potential field with the same photospheric boundary conditions thus releasing the maximum amount of magnetic energy. Specifically, it delivers an energy  of the order of $\sim (0.7-1.2) B_{0}^{2}R^{3}/\mu$, i.e., ranging from $3 \times 10^{26}$~J in the case of sun-like stars  to $\sim 10^{31}$~J in the case of young stars with surface fields of $\sim 350$~G, with the specific value depending on the field configuration.  This mechanism preferentially  operates in young stars with a closed corona extending beyond the periastron distance of their close-in planets because in that case the perturbation by the planet is maximum. 

If the energy and helicity of the field are not large enough to produce an eruption, the second mechanism can be relevant because it is based on a transition from a non-linear field to a minimum energy linear field with the same helicity. This mechanism can release a lower amount of energy than the previous one, up to $(0.3-0.8) B_{0}R^{3}/\mu$ for the typical separations of close-by planets (cf. Fig.~\ref{free_energy}). However, it is still  enough to account for the typical flare energy in late-type stars. Again, the planet may act as a trigger, but this mechanism can operate  also in stars without companions.  The closed linear field needs to have a sufficient radial extension to get an energy lower than that of the initial non-linear field with the same helicity. This favors stars with an intense photospheric field ($B_{0} \geq 100-300$~G) because their closed loops can extend up to several stellar radii. For this reason, young active stars with hot Jupiters, such as V830~Tau and TAP~26, are the ideal targets to search for the energetic flares induced by star-planet magnetic interaction through this mechanism or the previous one. Systems with eccentric orbits are the best potential candidates because the interaction is expected to be maximum near the periastron allowing us to discriminate it from the ordinary flare activity of the star not induced by the  planet. 

Finally, we considered a mechanism that operates in the open-field corona of weakly or moderately active stars. It is related to the energy perturbation produced by the planetary magnetosphere, that is, it cannot operate in stars without close-by planets.  In this mechanism, the energy released ranges between $\sim 0.002 B_{0}^{2} R^{3}/\mu$ and $\sim 0.1 B_{0}^{2} R^{3}/\mu$, depending on the distance of the hot Jupiter and the ratio of its magnetic field to the stellar field (cf. Fig.~\ref{free_energy_radial_field}). Therefore, our model can be used to estimate the planet magnetic field when  spectropolarimetric measurements of the stellar field are available. This is not possible with the previous two mechanisms where the planet magnetosphere acts simply as a trigger. 

This third model can account for the energy of the flares observed in HD~17156 and HD~189733. The energy is released when the planet moves across a radial structure of the inhomogeneous stellar magnetic field such as a tall coronal streamer. In the case of HD~80606, the predicted energy release  is about $4$ times that of HD~17156, if its surface field is $B_{0} \sim 10$~G, confirming that this system is worth of a systematic monitoring for flaring activity close to periastron as suggested by \citet{Figueiraetal16}. A measurement of its coronal activity level or photospheric magnetic field would be welcome to exclude a pathologically weak surface field as in the case of WASP-18 \citep{Pillitterietal14a} that could make its flares extremely weak, sporadic, and undetectable. Indeed, the low level of its chromospheric activity \citep[$\log R^{\prime}_{\rm HK} = -5.06$,][]{Figueiraetal16}, suggests that it is a rather inactive star. 

\begin{acknowledgements}
The author is grateful to an anonymous Referee for a careful reading of the manuscript and several suggestions that helped him to substantially improve the present work. AFL wishes to thank Drs. A.~Maggio, S.~Sciortino, I.~Pillitteri, I.~Pagano, and S.~Desidera for interesting discussions on magnetic  star-planet interactions. This work was partially supported by the project WOW, one of the {\it Progetti Premiali} of the Italian Ministry of Education, University, and Research funded to the Italian National Institute for Astrophysics (INAF).  
\end{acknowledgements}

\appendix
\section{Energy of the magnetospheric field}
\label{app1}
We consider the energy of the field inside the planetary magnetosphere by adopting the model of \citet{Voigt81}. For the sake of simplicity, we assume that $C_{\rm d} = C_{\rm imf} =0 $ because this leads to the maximum energy variation for the stellar coronal field (cf. Eq.~\ref{energy_var_magsp}). The case $C_{\rm imf} \not=0$ can be treated in a similar way, but the calculations are somewhat more complicated, so we focus on this simpler case. 

The magnetic field inside the magnetosphere ${\vec B}_{\rm int}$ is given by \citep[cf. Eqs. 3.14 and 3.22 in][]{Voigt81}:
\begin{equation}
{\vec B}_{\rm int} = {\vec B}_{\rm s} -\nabla u_{\rm cfs}, 
\end{equation}
where ${\vec B}_{\rm s}$ is the field produced by the planetary dipole and the current systems inside the  magnetosphere that we assume to be independent of the star-planet separation, while $u_{\rm cfs}$ is the potential of the inner Chapman-Ferraro field that  satisfies the boundary condition: $\partial u_{\rm cfs} / \partial n =  {\vec B}_{\rm s} \cdot \hat{\vec n}$ \citep[cf. Eq.~5.1 in][]{Voigt81}, where $\hat{\vec n}$ is the unit normal to the magnetopause. The energy $E_{\rm m}$ of the field inside the volume of the magnetosphere $V_{\rm m}$ is:
\begin{equation}
2 \mu E_{\rm m} = \int_{V_{\rm m}} ({\vec B}_{\rm s} - \nabla u_{\rm cfs})^{2} \ dV = \int_{V_{\rm m}} [B_{\rm s}^{2} - \nabla u_{\rm cfs} \cdot (2{\vec B}_{\rm s} - \nabla u_{\rm cfs} )] \, dV. 
\end{equation} 
Since $\nabla \cdot [(2 {\vec B}_{\rm s} - \nabla u_{\rm cfs}) u_{\rm cfs}] = \nabla u_{\rm cfs} \cdot (2 {\vec B}_{\rm s} - \nabla u_{\rm cfs})$ because ${\vec B}_{\rm s}$ is solenoidal and $u_{\rm cfs}$ satisfies the Laplace equation,  we can apply the divergence theorem and find: 
\begin{equation}
2 \mu E_{\rm m} = \int_{V_{\rm m}} B_{\rm s}^{2} \, dV - \int_{S(V_{\rm m})} u_{\rm cfs} ({\vec B}_{\rm s} \cdot \hat{\vec n}) \, dS,
\label{magsph_energy}
\end{equation}
where $S(V_{\rm m})$ is the surface of the magnetosphere and we made use of the boundary condition $({\vec B}_{\rm s} -\nabla u_{\rm cfs})\cdot \hat{\vec n} = 0$.  Making again use of the boundary condition:
\begin{equation}
2 \mu E_{\rm m} = \int_{V_{\rm m}} B_{\rm s}^{2} \, dV - \frac{1}{2}\int_{S(V_{\rm m})} \frac{\partial u_{\rm cfs}^{2}}{\partial n}  \, dS. 
\label{magsph_energy_fin}
\end{equation}
The linearity of the Laplace equation and the boundary condition imply that $u_{\rm cfs}$ and $\partial u_{\rm cfs} / \partial n$ are directly proportional to the field strength $B_{\rm s}$ on the magnetopause, yielding $\partial u_{\rm cfs}^{2}/ \partial n \propto B_{\rm s}^{2} > 0$  \citep[note that the derivative of the inner potential $u_{\rm cfs}$ along the outward normal $\hat{\vec n}$ has the same sign as $u_{\rm cfs}$; cf. Eqs. 5.11, 5.18, and 5.27 in][e.g., in the simple case of an aligned planetary field, $\chi = \psi = 0 \Rightarrow f_{2}(\chi, \psi) = 0$ in his Eq.~5.9]{Voigt81}.  
We can neglect the contribution of the planetary surface to the second integral in the r.h.s. of Eq.~(\ref{magsph_energy_fin}) because $u_{\rm cfs}$ is negligible there.
 Therefore, we see that the second term in the r.h.s. of Eq.~(\ref{magsph_energy_fin}) decreases when the planet comes closer to the star because the radius $R_{\rm m}$ of the magnetosphere decreases (cf. Eq.~\ref{rm_eq}) and $B_{\rm s}(R_{\rm m}) \propto R_{\rm m}^{-\gamma}$, where $\gamma \geq 2$. Also the first integral in the r.h.s. of Eq.~(\ref{magsph_energy_fin}) decreases because the volume of the magnetosphere $V_{\rm m}$ decreases. Since  Eq.~(\ref{energy_var_magsp}) gives a greater negative energy value when the planet is closer to the star, we conclude that the total magnetic energy of the star-planet system (i.e., the energy of the stellar coronal field plus the energy of the magnetospheric field) decreases when the planet moves closer to the star.

\end{document}